\documentclass{jdsart}

\volume{0}
\issue{0}
\pubyear{2025}
\articletype{research-article}
\doi{0000}
\usepackage{caption}
\usepackage{subcaption}
\usepackage{natbib}
\usepackage{orcidlink}
\usepackage{float, latexsym, times}
\usepackage{amsfonts, amstext}
\usepackage{multirow}
\usepackage{booktabs}
\usepackage{comment}
\usepackage{color}
\usepackage{xcolor}

\usepackage{tabularx}

\begin{document}

\title{Perceptions and Utilization of GenAI Tools among Data Science Students and Faculty} 
\author{\inits{A.M.}\fnms{Abeer M.} \snm{Hasan}\thanksref{c1}\ead{amhasan1@ncat.edu}}
\author{\inits{S.A.}\fnms{Sayed A.} \snm{Mostafa}}
\thankstext[type=corresp,id=c1]{Corresponding author.}
\address{\institution{Department of Mathematics \& Statistics, North Carolina A\&T State University}, \cny{Greensboro, NC, USA}}


\setcounter{equation}{0} \numberwithin{equation}{section}
\date{}
\maketitle

\section*{Abstract}
This study investigates perceptions and use of generative artificial intelligence (GenAI) tools among students and faculty in statistics and data science at a historically Black college or university. Survey data from 119 valid student responses and 14 faculty responses were used to examine familiarity, usage patterns, perceived benefits, awareness of limitations, and instructional support needs. Students reported substantial use of GenAI, with ChatGPT as the dominant tool, primarily for coding assistance and writing support. Although student perceptions of AI in data science workflows and careers were generally positive, confidence in interpreting AI-generated outputs was limited, and concerns about accuracy, reliability, and over-reliance were common. Faculty also viewed GenAI favorably, but self-rated proficiency and the frequency of classroom integration remained limited. Comparisons across student subgroups suggested that familiarity with GenAI and awareness of its limitations varied more by academic level than by gender. These findings highlight a gap between AI adoption and AI literacy and underscore the need for structured training, validation practices, and clearer institutional guidance for responsible AI integration in data science education.\\


\noindent
Keywords: AI literacy; Data Science (DS) education, Generative AI (GenAI); Limitations; Perceptions.

\section{Introduction \& Background}\label{sec:Lit}
Generative AI refers to a class of models that can create new data based on patterns and structures learned from existing data. These models rely on deep learning techniques and neural networks to generate content across domains such as text, images, and code. Since OpenAI released ChatGPT at the end of 2022, a competitive landscape of powerful tools, including Claude, Gemini, and Copilot, has emerged as a disruptive force across industries \citep{michel2023challenges}. 

The field of data science (DS) is central to this transformation. GenAI tools are reshaping professional workflows by automating complex tasks in data wrangling, visualization, and analysis, thereby accelerating insight generation and fostering actionable intelligence.
This technological shift has profound implications for the DS workforce and, consequently, for higher education. Early evidence indicates that AI is already altering the job market, depressing early-career employment in AI-exposed occupations \citep{brynjolfsson2025canaries}. This reality creates an urgent imperative for universities to adapt curricula to equip students with the skills to navigate and leverage this new AI-augmented ecosystem.

A growing body of literature suggests that integrating GenAI can significantly enhance the learning experience in DS. AI tools can automate repetitive tasks such as writing computer code, allowing students to focus on conceptual understanding and critical thinking \citep{holmes2023guidance}. Studies suggest that GenAI-driven educational tools can improve academic performance \citep{jauhiainen2023generative}.
Beyond simple automation, AI is being positioned as a pedagogical partner. For example, AI-generated adaptive feedback has been shown to improve students' statistical skills \citep{bauer2025effects}.

Despite clear opportunities, the literature extensively documents formidable barriers to integrating GenAI. Perceptions among educators vary widely, ranging from enthusiasm for innovation to skepticism about the ethical implications and integrity of course assessments \citep{prilop2025generative}. A primary concern is the blurring of the line around academic misconduct, as the ease of content generation complicates traditional definitions of plagiarism and necessitates rethinking assessment strategies \citep{duah2024generative}.

Students' reliance on AI to complete assignments complicates the evaluation of their true understanding and effort. 
\cite{10.1145/3585059.3611449}  discuss the potential consequences of AI-assisted cheating, including risks to educational quality, fairness, and the credibility of academic institutions. While \citep{wang_2025} focus on the limitations of AI detection tools, the inadequacies of traditional teaching methods in this context, and the potential for the responsible integration of AI into educational practices. 
Significant ethical concerns further complicate the adoption. GenAI models can perpetuate and amplify biases present in their training data, raising questions about equity in educational content \citep{gupta2025generative}. The security and privacy of student data also remain key institutional issues \citep{ko2025framework}.

These multifaceted challenges underscore that the successful integration of GenAI tools into DS is not merely a technical task but a complex socio-technical problem. It requires a coordinated effort to address faculty and student needs, establish clear ethical frameworks, and redesign curricula to balance the benefits of AI with the development of core competencies.

Our study provides empirical data on the student and faculty perspectives in DS education.
Understanding the perspectives of DS students and faculty is crucial for navigating this transition; however, research on their specific use and needs remains relatively limited \citep{chan2023students}. This paper aims to fill this gap by analyzing survey data from DS students and faculty regarding the integration of GenAI tools into their curricula.

\subsection*{Contribution and Study Relevance}
Existing research has largely emphasized broad perceptions of AI in education or the technical capacities of AI systems to support analysis, leaving a gap in understanding how AI tools are perceived and integrated within data-intensive STEM disciplines. 
A recent American Statistical Association (ASA) member survey provided an initial snapshot of how statisticians in academia and industry are engaging with generative AI, highlighting emerging use cases alongside persistent concerns about reliability, bias, ethics, and governance \citep{Glickman03042025}.
However, the descriptive nature of this effort underscores the need for more systematic, data-driven studies that rigorously examine patterns of use,  perceptions, awareness of limitations, and educational implications across stakeholder groups, especially DS students and educators.

This study offers a discipline-specific perspective by focusing on students and faculty at a Historically Black College or University (HBCU), where questions of fairness, trustworthiness, and responsible AI adoption are closely linked to broader concerns of equity and representation in STEM fields. This context provides an important perspective on responsible and equitable AI integration in technical education. 

In addition to the discipline-specific questions, we developed parallel survey instruments for students and faculty. By integrating input from both students and educators, we present a comprehensive understanding of the experience and needs of both groups. The findings aim to inform the responsible and equitable integration of GenAI in DS education.

Our study aims to address the following research questions: 
\begin{enumerate}[label=\textbf{RQ\arabic*}, leftmargin=2.5em]
\item \label{rq1} How familiar are students with GenAI tools for DS?
\item \label{rq2} What are the students' perceptions of the role of GenAI tools in DS education and workflow? 
\item \label{rq3} To what extent are students aware of the limitations and challenges of GenAI technology?
\item \label{rq4} How do gender or academic class levels influence students' overall familiarity, perceptions of GenAI usage, and awareness of limitations in DS?

\item \label{rq5} What are the faculty perceptions, competence, and concerns regarding integrating GenAI tools into DS education and workflow? 
\item \label{rq6} What are the institutional support and training needs of the faculty to effectively integrate GenAI literacy and tools into their DS courses?\\
\end{enumerate}

The remainder of the article is organized as follows.
Section \ref{sec:Methodology} discusses the study methodology.  
 {Section \ref{sec:Findings_Students} discusses findings for Questions 6--7, 10--14, 18--19, 21, and 24 of the student survey. Section \ref{sec:Findings_Faculty} presents findings from Questions 7--20, and 22--23 of the faculty survey. Additional descriptive results, including detailed response distributions for individual survey items, are reported in Section S1 of the Supplementary Materials.  
    Section \ref{sec:Discussion} presents study limitations, conclusions, and recommendations.
    Sections S2 and S3 of the Supplemental Materials include the complete survey instrument for students and faculty, respectively.}

\section{Methodology}\label{sec:Methodology}
 
This study was designed to evaluate the perceptions, adoption patterns, and principal concerns regarding AI tools among students and faculty in statistics and DS. To obtain a comprehensive view, we developed and administered two versions of the survey: one for students currently enrolled in a statistics or DS course and one for faculty who teach these courses. Participants were recruited via email invitations and flyers posted in the departments responsible for these courses. Data collection occurred during the academic year 2024-2025.

The survey instruments were adapted and expanded from the work of \cite{chan2023students} to specifically address data science curricula, workflows, and tools. We expanded the survey instrument to include 24 questions for students and 23 questions for faculty.  Using a combination of multiple-choice and Likert scale questions, the surveys evaluated familiarity with AI tools, common use cases, perceived benefits and risks, awareness of institutional policies, and training needs for faculty and students. 
The responses were analyzed using descriptive statistics and thematic categorization.

 We constructed three composite scores reflecting student familiarity with AI tools, perceptions of AI integration in data science, and awareness of AI limitations. Confirmatory factor analysis was used to evaluate the internal structure. Welch's t-test was used for 2-group comparisons (Males vs. females), and ANOVA with Tukey post hoc was used for 3-group comparisons (lower-division, upper-division, and graduate students).

 The survey was administered through Qualtrics. Data analysis and visualizations were performed using R version 4.4.2 \citep{RCoreTeam}. 
\textcolor{black}{The complete student and faculty survey questions are provided in Sections S2 and S3 of the supplementary materials. To ensure reproducibility, we provided complete anonymized data from both surveys and R code with question-by-question data processing, summarization, and visualizations in our project's GitHub repository (\url{https://github.com/Nothgisrandom/AI4DS}). 
}

\section{Results from the Student Survey} \label{sec:Findings_Students}


 We defined a valid response as one in which the participant spent at least 120 seconds and answered at least 25\% of the questions. Filtering out invalid responses yielded a sample of 119 students. The respondents could skip any question they did not feel comfortable answering. The number of valid (nonempty) responses varied across questions, and some questions allowed multiple selections. When applicable, we used the percentage of valid responses as the denominator to calculate relative frequencies and indicated the number of valid responses in each relevant figure. 
 
 The gender breakdown of the student respondents was: 71 female (61.2\%), 40 male (34.5\%), and 5 in other reported categories. The class distribution of the student participants was 28 (25.2\%) lower-division students (freshmen or sophomores), 57 (51.4\%) upper-division students (juniors or seniors), and 26 (23.4\%) graduate students. Most of the declared majors of student participants were DS, computer science/IT, or statistics. Freshmen were underrepresented in our survey because of the prerequisite sequence for our DS courses.

\subsection{Familiarity with GenAI Tools for DS}
To explore students' familiarity with AI tools for DS (\ref{rq1}), we used five survey questions. The majority of students reported at least some engagement with AI tools in their coursework: occasional use was the most common response (36.2\%), followed by frequent (20.7\%) and very frequent use (15.5\%). In contrast, 15.5\% had never used AI tools, and 12.1\% used them rarely. Students primarily engaged with AI for coding assistance (69.4\%) and proofreading or writing support (52.3\%), with more specialized applications such as mathematical proofs and literature reviews reported less frequently. These results indicate that students use AI tools predominantly for implementation support and refinement rather than for content generation.

Among specific tools, \emph{ChatGPT} was the most widely used (80\%), followed by \emph{Grammarly} (54\%). Familiarity with domain-specific platforms such as \emph{Google Colab} and \emph{AI-driven analytics tools} was notably lower, suggesting that students' current exposure is concentrated in general-purpose AI assistants rather than specialized data analytics tools.

\begin{figure}
    \centering
    \includegraphics[width=0.95\textwidth]{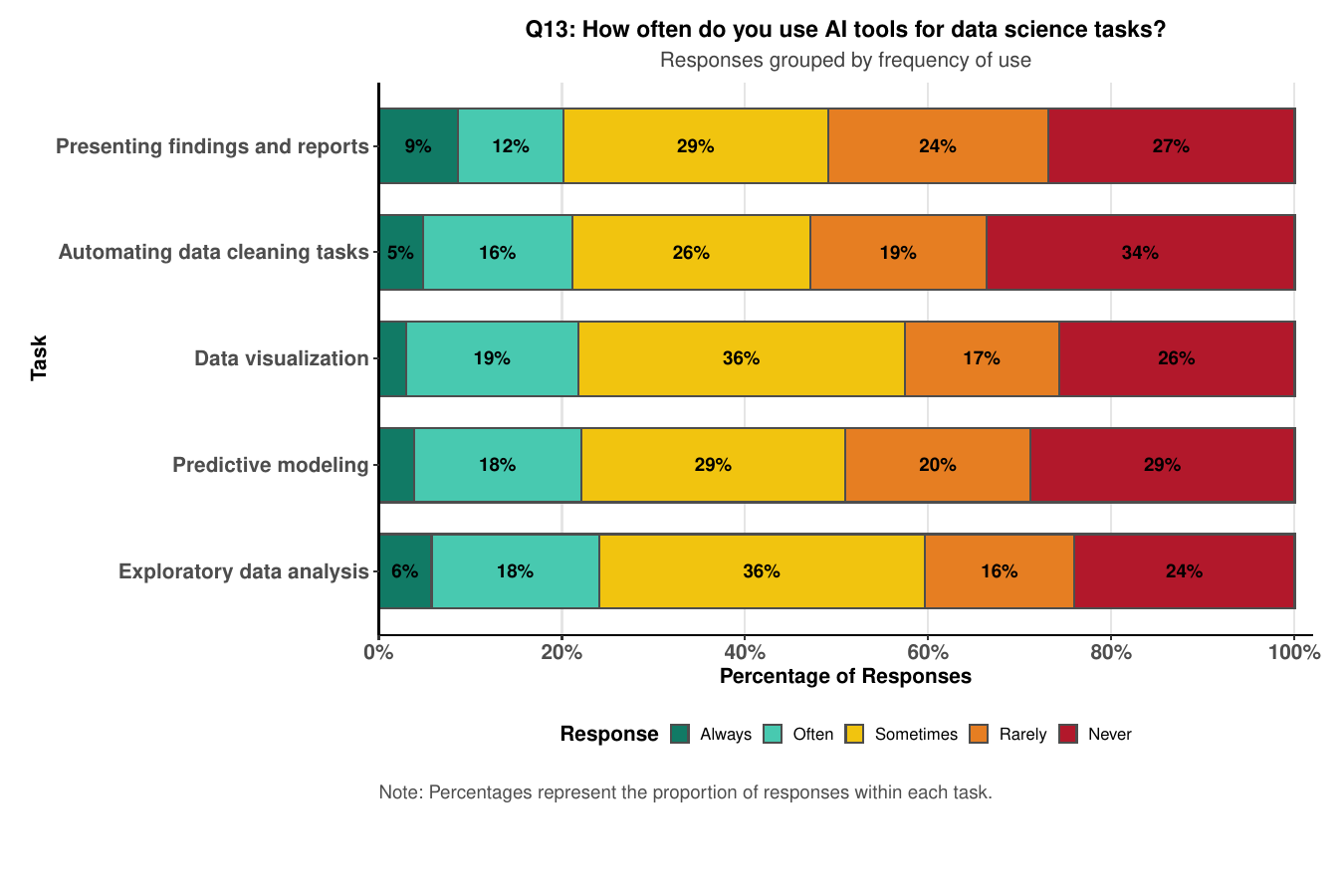}
    \caption{Students' self-reported usage frequency of AI tools in specific  DS  tasks.}
    \label{fig:Q13_utilization_horizontal}
\end{figure}

We also asked students to rate how often they use AI tools for specific DS tasks, including data wrangling and visualization, predictive modeling, exploratory data analysis, and the presentation of findings. Figure~\ref{fig:Q13_utilization_horizontal} summarizes the responses. Across all five task domains, the proportion of frequent users remained around 20\%, while a substantial share reported rare or no use. These patterns may reflect limited training on DS-specific tools, restrictions imposed by faculty, and potential underreporting due to students' reluctance to disclose AI use.{Detailed tool-by-tool familiarity ratings and task-level usage distributions are provided in Section~S1.1 of the Supplementary Materials.}
 
\subsection{Perceptions of GenAI Tools in DS}
We used a set of Likert-scale questions to assess students' perceptions of AI tools in DS (\ref{rq2}). Survey responses indicate that most students viewed AI tools favorably; they seemed optimistic about AI's potential to automate processes, improve accuracy, and generate new insights. Nearly 70\% of student respondents recognize the potential of AI tools to enhance accuracy, provide new insights, and automate data processing (Q12). 

\begin{figure}
    \centering
    \includegraphics[width=0.95\textwidth]{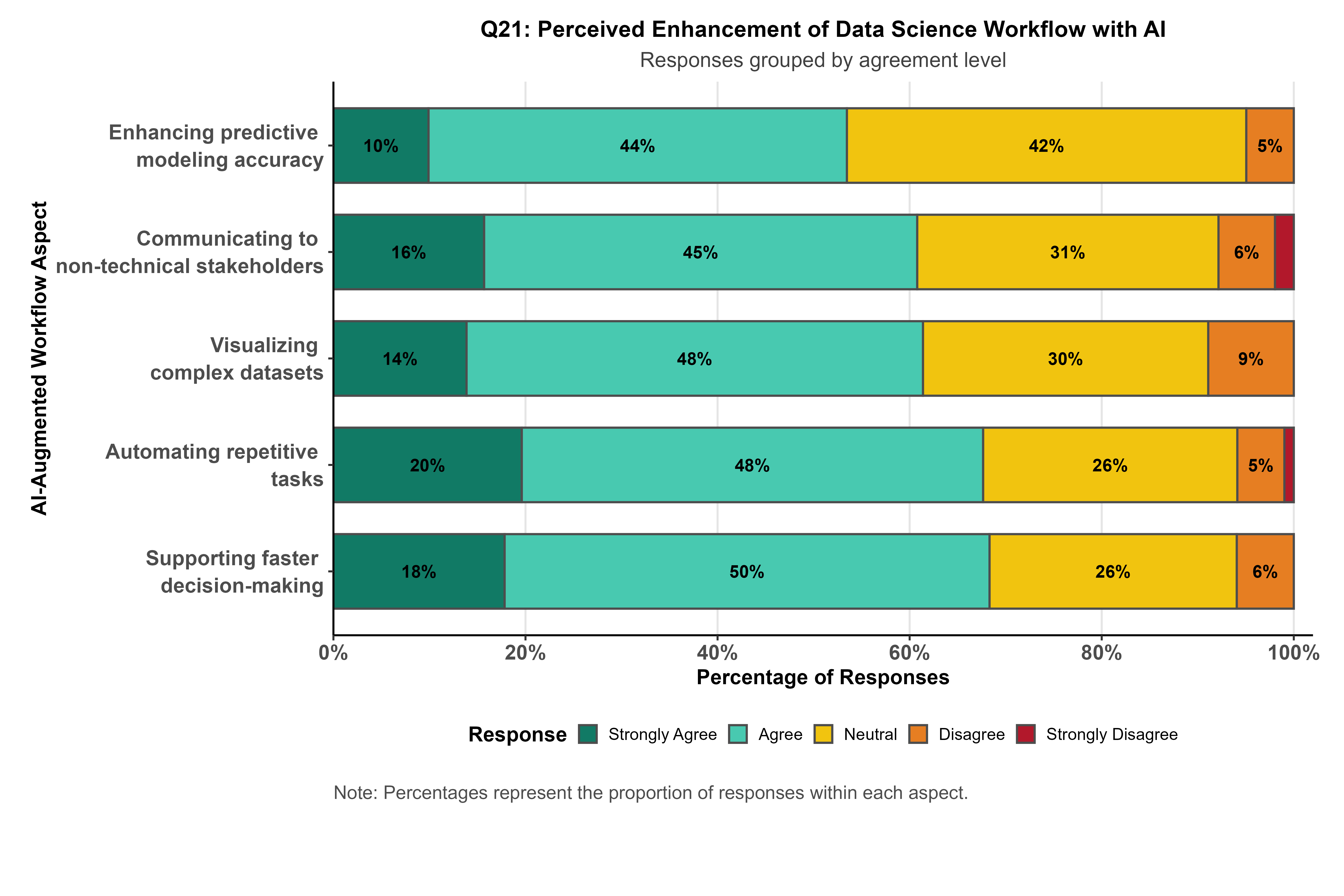}
      \caption{Students' perspective on AI's potential to enhance the DS workflow.}
    \label{fig:enhance_DSworkflow}
\end{figure}

Figure~\ref{fig:enhance_DSworkflow} presents students' perceptions of AI's potential to enhance the DS workflow (Q21). Across all five workflow dimensions, agreement substantially outweighed neutrality and disagreement. The strongest endorsements appeared for supporting faster decision-making and automating repetitive tasks, while perceptions regarding predictive modeling accuracy showed relatively more neutrality. This pattern suggests that students associate AI primarily with efficiency gains while recognizing that modeling performance depends on statistical rigor and data quality.

\textcolor{black}{Students also recognized the importance of AI for DS careers, when asked ``Do you think students should learn how to use AI tools for their data science careers?'', (Q16): 90.2\% recommended learning AI tools for DS careers and considered AI an essential (32.4\%) or a nice-to-have skill (57.8\%), and 72.4\% expressed willingness to incorporate AI tools into future DS projects (Q17). When asked whether AI tools should complement or replace traditional methods (Q24), a majority (54.9\%) favored a balance between AI tools and traditional methods, while 31.4\% favored using AI tools to complement traditional methods. Less than 8\% supported replacement.
These findings suggest a broad consensus favoring integrative rather than substitutive adoption of AI in DS. Complete response distributions for individual perception items are provided in Section~S1.2 of the Supplementary Materials.}

\subsection{Awareness of GenAI Limitations}
\begin{figure}
    \centering
    \includegraphics[width=0.85\textwidth]{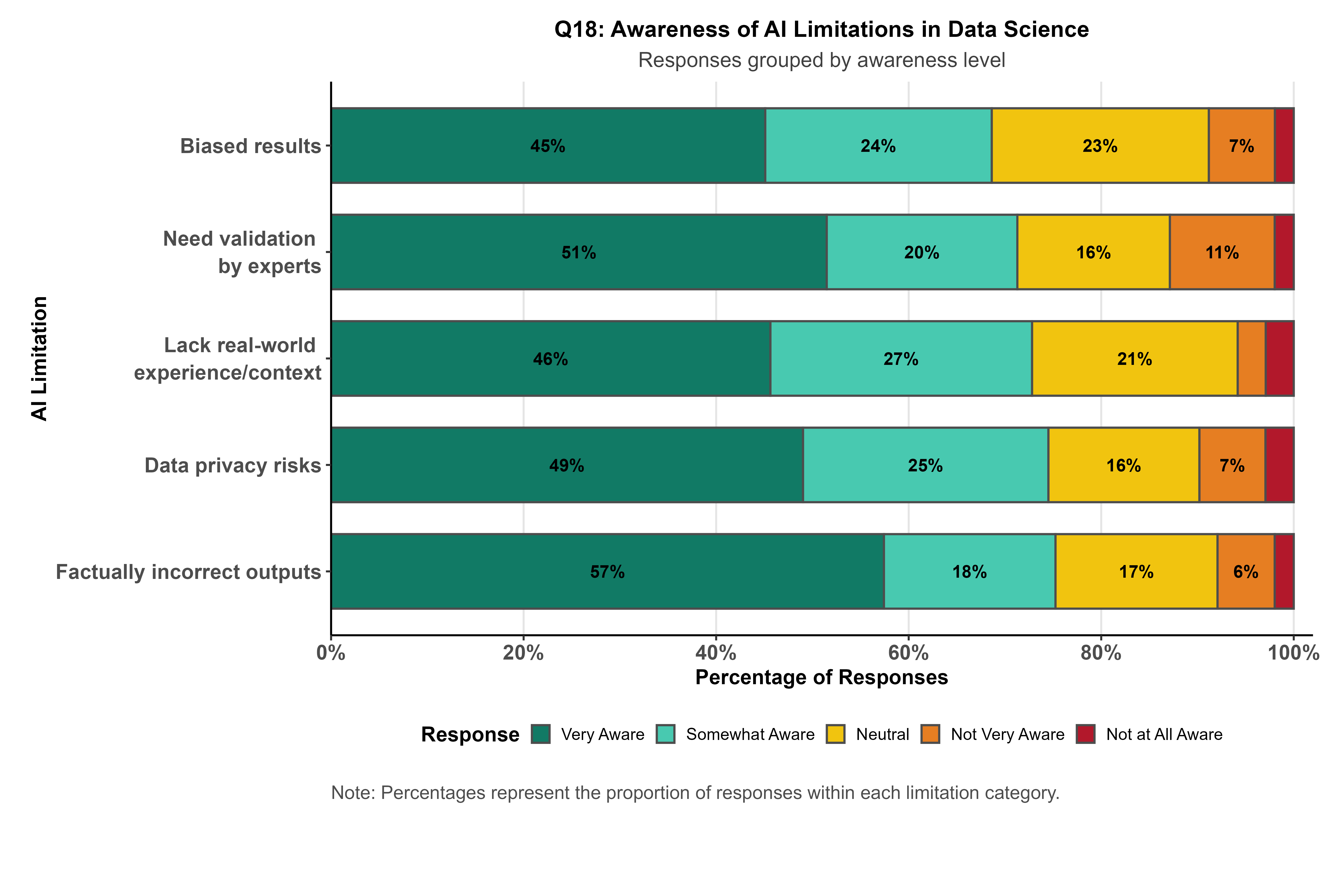}
    \caption{Students' awareness of the listed AI limitations in DS workflows.}
    \label{fig:Q18_Limitation}
\end{figure}

\textcolor{black}{To explore students' awareness of AI limitations in DS tasks (\ref{rq3}), we used three sets of Likert-scale questions (Q18).
More than 70\% of students recognized the limitations of AI tools, particularly the risks of inaccurate or biased outputs and the need for expert validation. When asked whether they had encountered issues such as hallucinations, bias, or inaccuracy (Q20), nearly 70\% reported experiencing such issues occasionally or frequently. Despite these encounters, students maintained an overall positive attitude toward AI's potential. Figure~\ref{fig:Q18_Limitation} summarizes students' awareness of specific AI limitations in DS workflows.}

Students' confidence in explaining AI-generated outputs was notably limited (Q14): only 40.6\% reported confidence, while 59.4\% were either not confident or neutral. This gap between usage and interpretive confidence underscores the need for instruction in output validation and effective prompting strategies.
Among students' primary concerns regarding AI in DS (Q19), accuracy and reliability were cited most frequently (23.6\%), followed by privacy and security, skill degradation, and ethical concerns or bias. In the educational context (Q23), over-reliance on AI emerged as the dominant concern (52.4\%), followed by hindering skill development (29.3\%). 
These priorities indicate that epistemic risk-'can I trust the output?'-appears to outweigh occupational risk in students' current assessment of AI tools. Additional details on students' ranking of their concerns and challenges with AI integration into DS are provided in Section~S1.3 of the Supplementary Materials.


\subsection{Gender and Class Level Differences}
\textcolor{black}{In this section, we examine whether gender or academic class level influences students' familiarity with GenAI tools, perceptions of their role, and awareness of their limitations in DS (\ref{rq4}). To facilitate these comparisons, we constructed three composite scores from the survey items, validated each score using confirmatory factor analysis, and tested for group differences using Welch's t-test and one-way ANOVA.}

\subsubsection{\textcolor{black}{Construction of Composite Scores}}

\textcolor{black}{We constructed three composite scores by aggregating responses from related sets of Likert-scale survey items, as detailed in Table~\ref{tab:ai_composite_items}. For each item, responses were coded on a numeric scale from 1 to 5, with higher values indicating greater familiarity, more favorable perceptions, or greater awareness of limitations. For example, in Question~10, ``Not at all Familiar'' was coded as 1 and ``Very Familiar'' as 5; in Question~18, ``Not at All Aware'' was coded as 1 and ``Very Aware'' as 5. Each composite score was computed as the arithmetic mean of the coded item responses within its respective construct, yielding a score that ranges from 1 to 5. The three constructs are:}
\textcolor{black}{\begin{itemize}
    \item \textbf{Familiarity with GenAI tools for DS}, derived from the five items in Question~10 (self-reported familiarity with specific AI tools);
    \item \textbf{Perceptions of GenAI use in DS}, derived from four items in Question~12 and five items in Question~21 (perceptions of AI's role in DS workflows);
    \item \textbf{Awareness of GenAI limitations}, derived from the five items in Question~18 (recognition of risks such as bias, hallucination, and privacy).
\end{itemize}}

\begin{table}[htbp]
\centering
\caption{Survey Items Used to Construct Composite Measures}
\label{tab:ai_composite_items}
\footnotesize
\setlength{\tabcolsep}{6pt}
\renewcommand{\arraystretch}{1.02}

\begin{tabularx}{\textwidth}{@{}X|X@{}}
\toprule
\begin{minipage}[t]{\linewidth}
\textbf{Familiarity with GenAI tools for DS}\\[0.3em]
\emph{Q10: How familiar are you with the following AI technologies in DS?}
\begin{itemize}
    \item ChatGPT
    \item Gemini
    \item Google Colab
    \item AI-powered data visualization tools (e.g., Tableau, Power BI with AI)
    \item AI-driven analytics platforms\\
\end{itemize}


\vspace{0.4em}
\textbf{Awareness of Limitations Score}\\[0.3em]
\emph{Q18: To what extent are you aware of the following limitations of using AI for DS tasks?}
\begin{itemize}
    \item AI tools can produce biased results due to biased training data.
    \item AI tools may generate factually incorrect outputs (hallucinations).
    \item AI models lack real-world experience and context understanding.
    \item AI-generated results often need to be validated by experts.
    \item Using AI tools could pose data privacy risks.
\end{itemize}
\end{minipage}
&
\begin{minipage}[t]{\linewidth}
\textbf{Perceptions of AI Tools in DS (Q12 and Q21)}\\[0.4em]
\emph{Q12: To what extent do you agree with the following statements regarding AI tools in DS?}
\begin{itemize}
    \item AI tools can enhance the accuracy of DS models.
    \item AI tools can automate tedious data processing tasks.
    \item AI tools can provide new insights or perspectives I would not have discovered otherwise.
    \item AI tools lack emotional intelligence, which may limit their utility in complex tasks.
\end{itemize}

\vspace{0.4em}
\emph{Q21: Do you think AI tools can significantly enhance the following aspects of your DS workflow?}
\begin{itemize}
    \item Automating repetitive data processing tasks.
    \item Enhancing predictive modeling accuracy.
    \item Visualizing complex datasets more effectively.
    \item Supporting faster decision-making.
    \item Helping communicate findings to non-technical stakeholders.
\end{itemize}
\end{minipage}\\
\bottomrule
\end{tabularx}
\end{table}

\subsubsection{\textcolor{black}{Confirmatory Factor Analysis}}
\textcolor{black}{Before using these composite scores in group comparisons, we conducted confirmatory factor analysis (CFA) to evaluate whether the items within each construct adequately reflect a single underlying latent factor. CFA assesses model fit by comparing the covariance structure implied by the hypothesized factor model to the observed covariance structure in the data. We report the following fit indices, which are standard in applied survey research:}

\textcolor{black}{\begin{itemize}
    \item The \textbf{chi-square statistic} ($\chi^2$) tests the null hypothesis that the model fits the data exactly. A non-significant $p$-value ($p > 0.05$) suggests an acceptable fit, although this test is sensitive to sample size.
    \item The \textbf{Comparative Fit Index (CFI)} and \textbf{Tucker-Lewis Index (TLI)} are incremental fit indices that compare the specified model to a baseline (independence) model. Values of 0.95 or above indicate a good fit \citep{hu1999cutoff}.
    \item The \textbf{Root Mean Square Error of Approximation (RMSEA)} quantifies the discrepancy between the model and the data per degree of freedom. Values below 0.06 indicate a close fit, while values between 0.06 and 0.08 reflect an acceptable fit. However, RMSEA is known to be upwardly biased in models with small degrees of freedom \citep{kenny2015performance}, which is relevant for the five-item constructs examined here.
    \item \textbf{Cronbach's alpha} measures internal consistency, with values above 0.70 generally considered acceptable for survey research. We report both raw and standardized alpha values; close agreement between them indicates stable reliability across items.\\
\end{itemize}}

 For the \textit{familiarity} construct (Q10), the model demonstrated adequate fit, $\chi^2(5) = 9.78$, $p = 0.082$, with a CFI of 0.96, TLI of 0.92, and RMSEA of 0.096. The slightly elevated RMSEA is consistent with the known upward bias of this index in models with few degrees of freedom \citep{kenny2015performance}. The \textit{perceptions} construct (Q12, Q21) also showed acceptable fit, $\chi^2(5) = 8.39$, $p = 0.136$, with strong incremental indices (CFI = 0.97, TLI = 0.94) and RMSEA of 0.083. The \textit{awareness of limitations} construct (Q18) exhibited excellent fit across all indices: $\chi^2(5) = 3.75$, $p = 0.586$, CFI = 1.00, TLI = 1.01, and RMSEA $\approx$ 0. Cronbach's alpha coefficients ranged from 0.76 to 0.86 across the three constructs, with standardized values closely aligned, indicating acceptable internal consistency. 
 These results provide support for the internal consistency and unidimensional structure of the composite measures.
 Table~\ref{tab:CFA} summarizes the CFA results for the {Familiarity}, {Perceptions}, and {Limitations} constructs.
\begin{table}[H]
\footnotesize
\centering
\caption{Model fit statistics from confirmatory factor analysis.}
\label{tab:CFA}
\setlength{\tabcolsep}{4pt}
\begin{tabularx}{\textwidth}{l c c c c  c c }
\toprule
\textbf{Composite} & \textbf{$\chi^2$ Stat (p-value)} & \textbf{CFI} & \textbf{TLI} & \textbf{RMSEA} & \textbf{Raw Alpha} & \textbf{Standard Alpha}  \\
\midrule
{Familiarity} &9.780 (0.082) & 0.962 & 0.924 & 0.096 & 0.758 & 0.754\\
\addlinespace
{Perceptions}  & 8.389 (0.136) & 0.970 & 0.940 & 0.083 & 0.765 & 0.763\\
\addlinespace
{Awareness of Limitations}  & 3.750 (0.586) & 1.000 & 1.012 & 0.000 & 0.859 & 0.859\\
\bottomrule
\end{tabularx}
\end{table}

\subsubsection{\textcolor{black}{Hypotheses and Statistical Tests}}

\textcolor{black}{For each of the validated composite scores, we tested for statistically significant differences across gender and academic class level. For example, for the \textit{Familiarity} score, we formulated the hypotheses in equations \ref{eq:hyp_gender} and 
\ref{eq:hyp_level3}. Similar hypotheses were used for the \textit{Perceptions} and \textit{Limitations} scores. }

\begin{equation}
\label{eq:hyp_gender}
\left\{
\begin{aligned}
H_{01}:&\ \text{\textcolor{black}{There is no difference in \textit{familiarity scores} between male and female students.}} \\
H_{A1}:&\ \text{\textcolor{black}{There is a difference in \textit{familiarity scores} between male and female students. }} 
\end{aligned}
\right.
\end{equation}

\begin{equation}
\label{eq:hyp_level3}
\left\{
\begin{aligned}
H_{02}:&\ \text{\textcolor{black}{There is no difference in \textit{familiarity scores} among lower-division, upper-division,}} \\
       &\ \text{\textcolor{black}{ and graduate students.}} \\
H_{A2}:&\ \text{\textcolor{black}{At least one group differs in \textit{familiarity score} among lower-division,  upper-division,}} \\
       &\ \text{\textcolor{black}{ and graduate students.}}
\end{aligned}
\right.
\end{equation}

\textcolor{black}{All tests were evaluated at the overall $0.05$ significance level. For two-group comparisons (\ref{eq:hyp_gender}), we used Welch's two-sample t-test, which does not assume equal variances across groups and is robust for unequal sample sizes. Since we are comparing three composite scores for each subgroup, we adjusted our significance level to control the family-wise type I error rate using a Bonferroni correction, setting the adjusted significance level to 0.05/3 = 0.0167. 
For the three-group comparison (\ref{eq:hyp_level3}), we used one-way analysis of variance (ANOVA), followed by Tukey's post hoc pairwise comparisons where the omnibus test was significant.}

\subsubsection{\textcolor{black}{Results}}
Table~\ref{tab:demographic_comparisons} presents the composite score means, standard deviations, and $p$-values for each group comparison.

\paragraph{\textcolor{black}{Gender.}}{Across the three composite scores, males had a higher arithmetic mean than females.  Welch's t-tests revealed no statistically significant differences in \textit{Familiarity} ($p = 0.560$) or \textit{Perceptions} ($p = 0.197$) between female and male respondents. However, a statistically significant difference was observed for \textit{Awareness of AI Limitations} ($p = 0.001$), with male respondents reporting higher mean awareness scores than female respondents.}

\paragraph{\textcolor{black}{Lower-Division, Upper-Division, and Graduate.}}{Across all three constructs, mean scores increased with academic standing. ANOVA tests indicated significant differences among the three class levels for \textit{Familiarity} ($p = 0.002$) and \textit{Awareness of AI Limitations} ($p = 0.009$), but not for \textit{Perceptions} ($p = 0.993$).}

{Post hoc comparisons revealed that graduate students scored significantly higher than lower-division students on both \textit{Familiarity} ($p_{adj} = 0.001$) and \textit{Awareness} ($p_{adj} = 0.011$). The difference in \textit{Awareness} between lower-division and upper-division students had ($p_{adj} = 0.027$). Note that the post-hoc pairwise comparisons were already adjusted for multiplicity using Tukey's HSD's method.}

\textcolor{black}{Overall, these comparisons suggest that differences in familiarity with GenAI tools and awareness of their limitations are more strongly related to academic experience and progression than with gender. The relative stability of perception scores across all subgroups indicates broadly consistent perceptions of AI in DS education, despite variation in exposure and self-reported knowledge.}


\begin{table}
\footnotesize
\centering
\caption{Composite score comparisons by gender and class levels.}
\label{tab:demographic_comparisons}
\setlength{\tabcolsep}{4pt}

\begin{tabular}{@{}l
    c c c
    !{\hspace{6pt}\vrule width 0.6pt\hspace{6pt}}
    c c c
    !{\hspace{6pt}\vrule width 0.6pt\hspace{6pt}}
    c c c@{}}
\toprule
\textbf{Group}
& \multicolumn{3}{c}{\textbf{Familiarity}}
& \multicolumn{3}{c}{\textbf{Perceptions}}
& \multicolumn{3}{c}{\textbf{Awareness of limitations}} \\
\cmidrule(lr){2-4} \cmidrule(lr){5-7} \cmidrule(lr){8-10}
& \textbf{n} & \textbf{Mean (SD)} & \textbf{P-value}
& \textbf{n} & \textbf{Mean (SD)} & \textbf{P-value}
& \textbf{n} & \textbf{Mean (SD)} & \textbf{P-value} \\
\midrule
Female
& 69 & 2.695 (0.921) &
& 70 & 3.767 (0.597) &
& 65 & 3.920 (0.939) & \\
Male
& 37 & 2.808 (0.963) & 0.560
& 37 & 3.924 (0.592) & 0.197
& 36 & 4.450 (0.572) & \textbf{0.001*} \\
\midrule
Lower-Division
& 27 & 2.265 (0.683) &
& 27 & 3.811 (0.639) &
& 24 & 3.658 (0.966) & \\
Upper-Division
& 54 & 2.755 (0.914) &
& 54 & 3.822 (0.544) &
& 52 & 4.196 (0.818) & \\
Graduate
& 26 & 3.162 (1.001) & \textbf{0.002*}
& 26 & 3.831 (0.678) & 0.993
& 25 & 4.360 (0.709) & \textbf{0.009*} \\
\bottomrule
\end{tabular}

\vspace{2pt}
\footnotesize\textit{Note.} Pairwise comparisons were conducted using Welch's \textit{t}-test, and three-group comparisons using ANOVA. To control the family-wise type I error rate using a Bonferroni correction, set the adjusted significance level to 0.05/3 = 0.0167. 
\end{table}


\section{Results from the Faculty Survey} \label{sec:Findings_Faculty}
We administered the faculty survey to faculty members who teach statistics or DS courses at our university. The number of valid responses was $ n=14$, with a gender breakdown of 9 females (64.3\%) and 5 males (35.7\%). We achieved a high response rate with nearly all statistics and data science faculty at our college completing the survey. The primary training of the faculty respondents was statistics or DS. They range in teaching experience from 1-3 years (38.5\%) to over 10 years (23.1\%).{This section highlights the most interesting faculty findings; additional details are provided in Section S1.4-S1.6. The complete survey instrument, including item-level summaries and visualizations, is reported in Section~S3 of the Supplementary Materials, with accompanying interpretive context.}

\subsection{Faculty Integration of GenAI Tools in Teaching DS}
\textcolor{black}{To explore faculty familiarity with AI tools in teaching DS (\ref{rq5}), we asked whether faculty had incorporated AI tools into their teaching (Q7), and which tools they use (Q8). A majority (64.3\%) reported incorporating AI tools at least sometimes, while 21.4\% had never done so. All faculty respondents reported using AI learning assistants such as {ChatGPT}, followed by {AI-powered coding platforms} (64.3\%) and {data analysis or visualization tools} (42.9\%). Use of automated grading and project management tools was minimal, consistent with student responses indicating that {ChatGPT} was the most widely adopted GenAI tool.}

\begin{figure}
    \centering
    \includegraphics[width=0.85\textwidth]{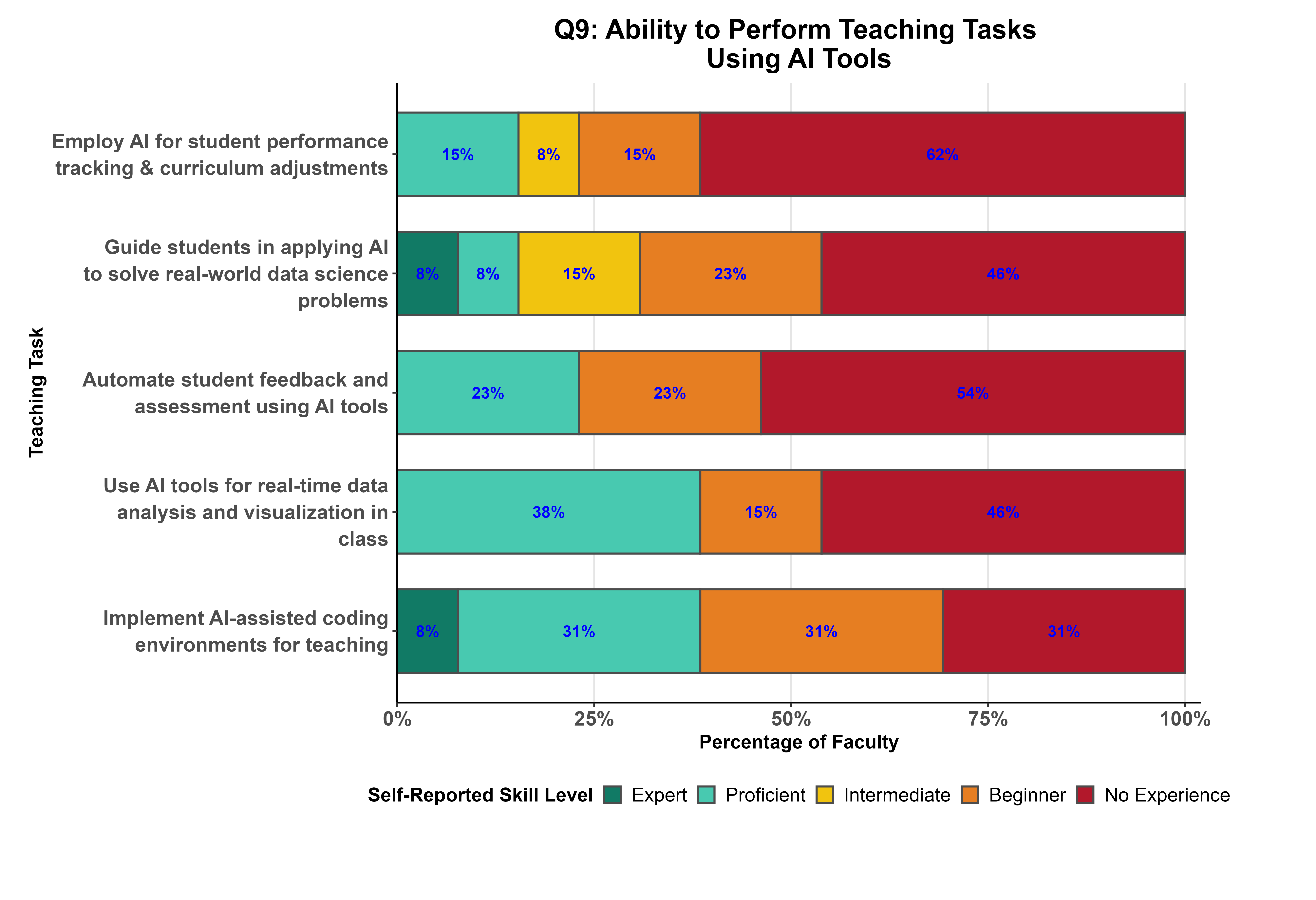}
    \caption{Faculty rating of their ability to perform teaching tasks using AI tools.}
    \label{fig:Fac_Ability}
\end{figure}

\begin{figure}
    \centering
    \includegraphics[width=0.85\textwidth]{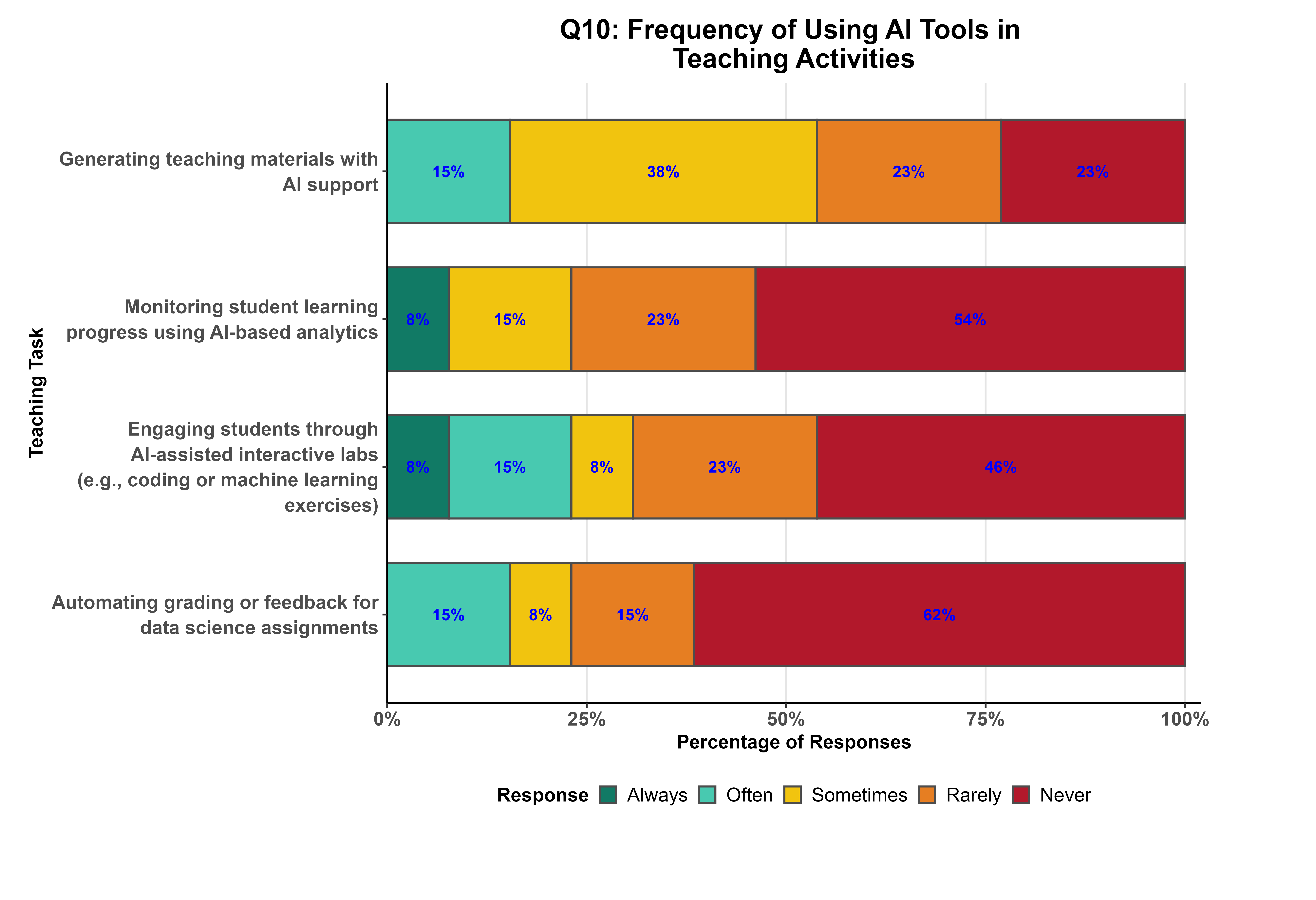}
    \caption{Faculty self-reported frequency of using AI tools in teaching activities.}
    \label{fig:Fac_usage}
\end{figure}

\textcolor{black}{When asked to rate their ability to perform teaching tasks using AI (Q9), over 60\% of faculty rated their skill level as ``Beginner'' or ``No Experience'' (Figure~\ref{fig:Fac_Ability}). This is a notable contrast to student self-assessments and may reflect either faster student adoption or differences in self-assessment across groups. The finding underscores the importance of professional development and training support as faculty integrate new technologies into DS education. Across four specific teaching activities (Q10), fewer than 23\% of faculty reported frequent AI use, and over 46\% reported never or rarely using AI for each activity (Figure~\ref{fig:Fac_usage}). The most common use cases were engaging students through AI-assisted interactive labs and generating teaching materials. Detailed distributions for faculty self-assessment and teaching activity usage are provided in Section~S1.4 of the Supplementary Materials.}

\subsection{Faculty Perceptions of Using AI Tools in DS}

We explored faculty perceptions of GenAI tools in teaching DS (\ref{rq5}). Overall, faculty expressed a generally positive view of AI adoption, coupled with nuanced concerns about its potential impact on student learning (Q11). Strong majorities agreed that AI tools can improve equity and inclusion in DS education (77\%), automate repetitive teaching tasks (nearly 70\%), and enhance instruction in technically complex topics such as machine learning and big data (over 75\%). A majority (77\%) also agreed that AI tools will be vital for the future of DS education. At the same time, faculty expressed notable concern that over-reliance on AI may reduce students' critical thinking skills, with responses split between agreement and neutrality rather than showing clear consensus. This tension, recognizing efficiency gains while guarding against erosion of deep learning, is a recurring theme in the results.

\begin{figure}[H]
    \centering
 \includegraphics[width=0.85\textwidth]{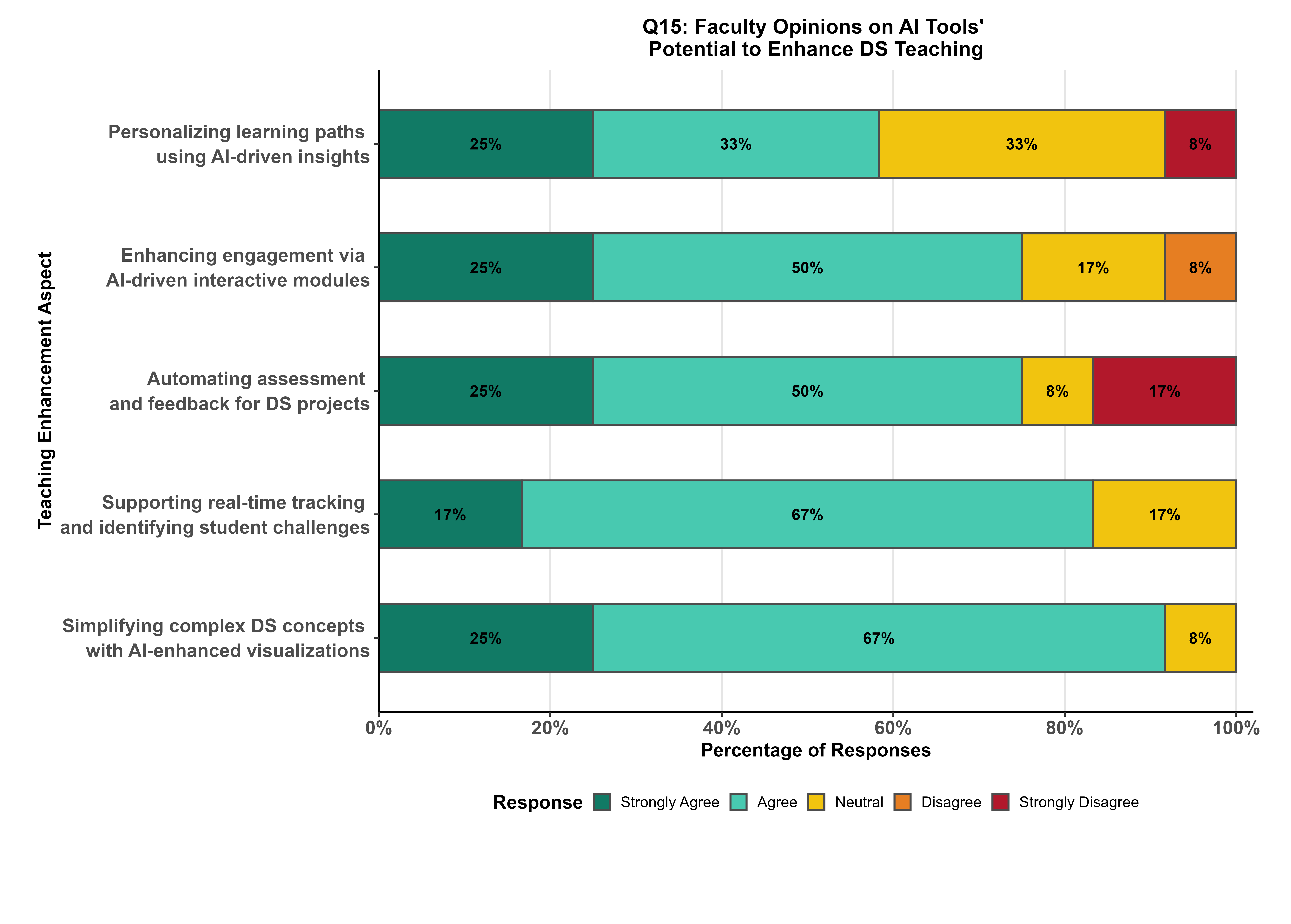}
    \caption{Faculty agreement on the potential for GenAI tools to improve DS education.}
  \label{fig:Fac_enhance_teaching_Q15}
\end{figure}

Figure~\ref{fig:Fac_enhance_teaching_Q15} presents faculty perceptions of AI as an enhancer of DS teaching (Q15). The strongest endorsement was for simplifying complex concepts through AI-enhanced visualizations, and no faculty disagreed with AI's potential for real-time tracking of student challenges. Perceptions were most mixed regarding personalizing learning paths, with the highest proportion of neutral and negative responses. These patterns suggest that faculty favor pedagogically grounded integration that preserves critical thinking and student agency. Additional detail on the equity and inclusion finding and per-item response distributions is provided in Section~S1.5 of the Supplementary Materials.


\subsection{Faculty Concerns Regarding AI Integration in DS Education}
\begin{figure}[H]
    \centering
    \includegraphics[width=0.85\textwidth]{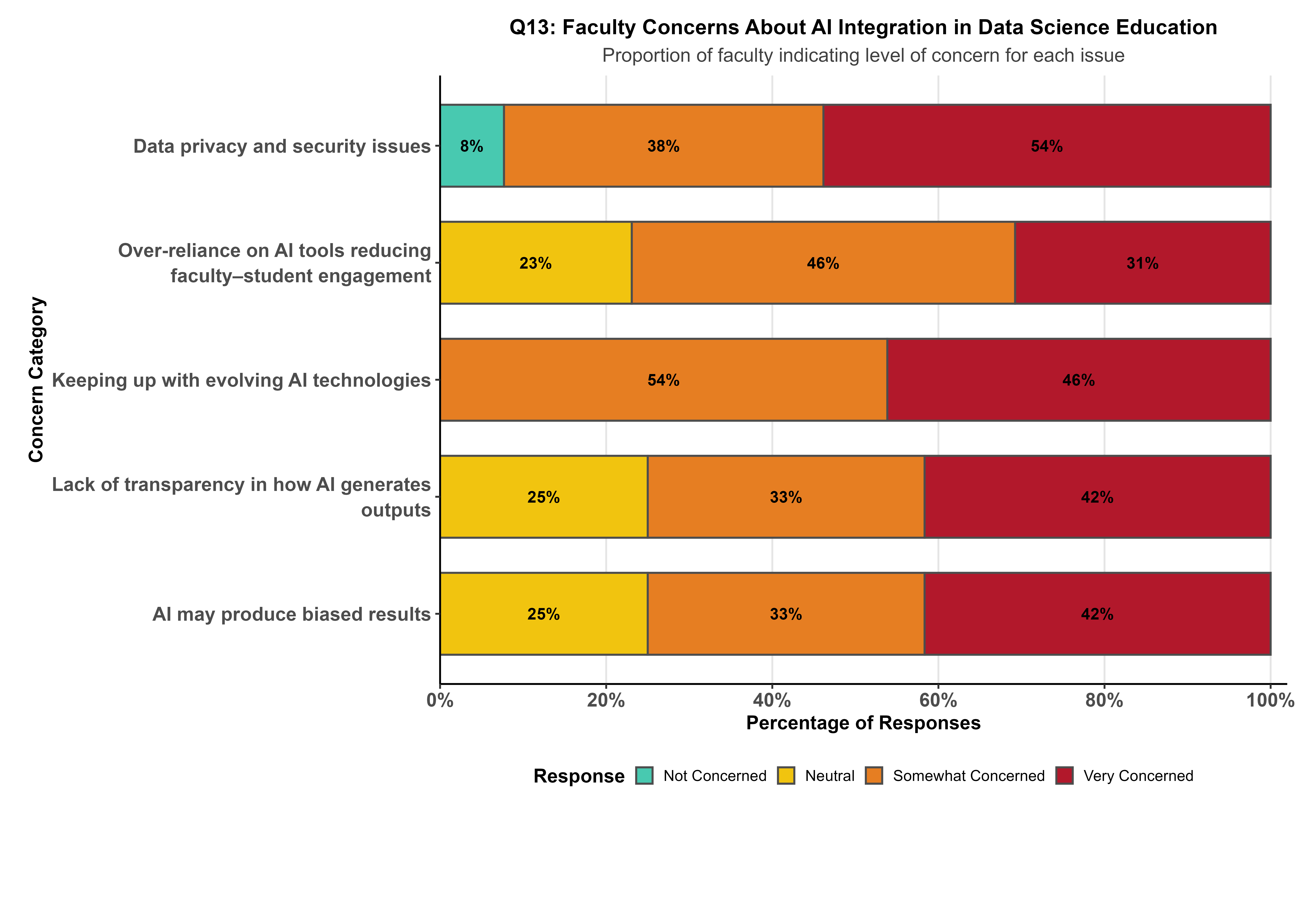}
    \caption{Faculty concerns regarding integrating GenAI tools into DS education.}
    \label{fig:Fac_ConcernsQ13}
\end{figure}

\textcolor{black}{We explored faculty concerns regarding AI tools in teaching DS (\ref{rq5}). Figure~\ref{fig:Fac_ConcernsQ13} summarizes the responses of Q13. Faculty were unanimous in their concern about keeping up with evolving AI technologies. Concerns about biased results in student assessments, impacting faculty-student engagement, and a lack of transparency in AI outputs were also widespread (at least 75\% each). Data privacy and security, while still a majority concern, were rated somewhat lower, with 8\% expressing no concern, which may reflect DS faculty's generally higher comfort with technology.}

\textcolor{black}{Regarding ethical priorities (Q18), 84.6\% of faculty cited excessive reliance on GenAI tools leading to reduced human oversight as their primary concern, and all respondents agreed that maintaining human supervision is essential. The most frequently cited integration challenges were insufficient institutional support or training, balancing AI with curriculum learning objectives, and ethical concerns such as fairness and bias.}

\textcolor{black}{Syllabus policies varied across respondents (Q20): 83\% allowed AI use for instructor-designated activities with required disclosure, while approximately 17\% prohibited AI entirely; none permitted unrestricted use. More than a third of faculty were unaware of institutional AI guidelines. This absence of uniform policy within the same department risks creating confusion for students regarding the boundaries between acceptable assistance and academic dishonesty.}

\textcolor{black}{To address these challenges (\ref{rq6}), faculty identified training and support needs centered on access to advanced AI platforms, technical integration support, and professional development opportunities (Q22). A majority (53.8\%) reported that their institution does not provide sufficient AI guidance or resources (Q23), highlighting a significant opportunity for institutional leadership. Complete response distributions for concerns, institutional awareness, and training needs are provided in Section~S1.6 of the Supplementary Materials.}

\section{Discussion} \label{sec:Discussion}
This study provides a descriptive quantitative account of generative AI adoption within a data science educational environment, explicitly comparing student and faculty populations. The results indicate that AI integration in DS education is neither uniform nor unidimensional. Rather, it reflects structured heterogeneity across usage intensity, interpretive confidence, and perceived risk.

From a measurement standpoint, the survey instrument operationalizes three latent dimensions: (i) familiarity with AI tools, (ii) awareness of limitations, and (iii) perceptions of GenAI tools in teaching statistics and DS content courses. Confirmatory factor analysis supports the multidimensional structure of these constructs. The moderate correlations among factors indicate related but distinct dimensions of AI engagement. Importantly, group comparisons demonstrate that AI engagement appears multidimensional and reflects interacting cognitive and attitudinal components.

Students report a significantly higher frequency of GenAI use than faculty (Figure \ref{fig:Q13_utilization_horizontal} vs. Figure \ref{fig:Fac_usage}). This adoption asymmetry suggests differential exposure and incentive structures across stakeholder groups. Students appear to integrate AI tools into operational workflows, whereas faculty engagement is more cautious and pedagogically constrained. The divergence is not merely behavioral; it is accompanied by distinct confidence profiles. Students report lower confidence in interpreting AI-generated outputs, while faculty report limited confidence in using AI to support teaching tasks. The coexistence of high utilization and low interpretive confidence among students raises concerns about the acceptance of unverified outputs.

A notable empirical finding is that both faculty and students rank concerns about over-reliance and skill erosion above privacy-related concerns. This prioritization suggests that perceived epistemic risk dominates technical risk in this context. In DS education, where statistical reasoning, inferential judgment, and model interpretation are core competencies, this distinction is particularly consequential. If AI systems function as probabilistic generators rather than deterministic solvers, then DS curricula must emphasize output verification, uncertainty quantification, and model interpretability.

The observed discrepancies between utilization and interpretive confidence indicate that AI adoption is outpacing the development of AI literacy. This gap underscores the importance of embedding formal instruction on the limitations of generative models, the risks of hallucination, and prompt design principles within DS coursework. Absent such instruction, students may rely on outputs without sufficient scrutiny, potentially weakening inferential rigor.

From a curricular modeling perspective, AI integration should be conceptualized as augmentation rather than substitution. Structured AI reflection assignments that require prompt disclosure, output validation, and justification of acceptance decisions may help preserve independent reasoning. Future research should empirically evaluate whether such interventions mitigate skill erosion through experimental or quasi-experimental designs.

More broadly, the results suggest that AI adoption in DS education is best understood as a socio-technical process involving competence acquisition, trust calibration, and institutional infrastructure. Access barriers, subscription costs, and limited professional development may contribute to uneven adoption patterns. Quantitative modeling of these structural predictors represents a natural extension of this work.

\subsection{Implications for Data Science Education}
The findings indicate that DS curricula must advance along two interconnected paths. Students need operational competence in using generative AI tools, including effective prompt construction, iterative refinement, and an understanding of system constraints. At the same time, programs must strengthen core statistical reasoning, particularly skills in model validation, uncertainty quantification, and the principled interpretation of results. Technical proficiency without conceptual depth risks superficial analysis.

The widespread availability of generative AI introduces a new pedagogical challenge. Students may mistake polished outputs for analytically sound conclusions. Traditional statistical software executes well-defined procedures, whereas generative systems produce responses through probabilistic processes that may not explicitly communicate assumptions or uncertainty. This distinction requires explicit instructional attention. Students should be trained to question outputs by examining assumptions, considering alternative modeling strategies, and identifying potential sources of error or instability.

{Metacognition} becomes central in this environment. Effective use of AI tools depends on a learner’s ability to monitor their own reasoning, recognize when cognitive shortcuts are being taken, and determine whether AI assistance is enhancing or replacing understanding. Structured reflective practices can support this development. For example, students may be required to document how prompts evolved, justify acceptance or rejection of generated results, or independently verify outputs using conventional statistical methods. These activities encourage deliberate engagement rather than passive reliance.

{Interpretability} should also be emphasized. DS education has increasingly prioritized transparent modeling and explainable methods. The same standards should apply to generative systems. Students must understand that AI-generated outputs do not inherently guarantee correctness, reproducibility, or traceability. Assignments that require cross-validation of generated analyses, replication of results, or critical comparison between AI-assisted and independently derived solutions can preserve methodological rigor.

Historically, statistical software reduced the need for manual computation and shifted emphasis toward modeling decisions and interpretation. Generative AI represents a further shift by automating portions of analytical reasoning. The appropriate educational response is neither prohibition nor unconditional acceptance. Instead, curricula should cultivate supervisory competence. Students must learn to interrogate and evaluate AI outputs while retaining responsibility for inference and judgment.

Intentional integration of metacognitive training and interpretability principles into data science coursework is therefore essential. Without structured guidance, frequent use of GenAI tools, combined with limited interpretive confidence, may weaken foundational reasoning skills. With thoughtful instructional design, however, generative AI can promote more substantive engagement with underlying model assumptions, uncertainty, and responsibility for analytic decisions.

\subsection{Limitations and Future Research}

This study is limited to a single R2 HBCU in the southeastern United States. While the institutional context provides meaningful insight, external validity is limited. Multi-institutional replication across R1 universities, liberal arts colleges, and international contexts would strengthen generalizability.
The reliance on self-reported survey data introduces potential response bias. Individuals with strong views regarding AI may have been more likely to participate. Additionally, concerns about academic integrity policies may have led some students to underreport their use of AI. 

The analysis presented here is primarily descriptive with group comparisons. Faculty findings are especially descriptive due to the small respondent pool. Future extensions could incorporate inferential modeling, including regression analysis to predict usage frequency and longitudinal designs to assess changes over time. 
Finally, experimental studies evaluating the causal impact of structured AI literacy interventions on statistical reasoning outcomes represent an important next step. Such work would move beyond measurement of perception toward assessment of educational effectiveness. 

\begin{funding}
The study is funded in part by a National Science Foundation Grant No. EES 2510214. 
\end{funding}

\bibliographystyle{jds}  
\bibliography{Refs_Clean20} 
 \end{document}


\title{Perceptions and Utilization of GenAI Tools among Data Science Students and Faculty}
\subtitle{Supplementary Material}
\author{\inits{A.M.}\fnms{Abeer M.} \snm{Hasan}\thanksref{c1}\ead{amhasan1@ncat.edu}}
\author{\inits{S.A.}\fnms{Sayed A.} \snm{Mostafa}}
\thankstext[type=corresp,id=c1]{Corresponding author.}
\address{\institution{Department of Mathematics \& Statistics, North Carolina A\&T State University}, \cny{Greensboro, NC, USA}}

\numberwithin{equation}{section}
\date{}
\maketitle

This supplementary material contains (S1) additional findings from both the student and faculty surveys; (S2)  the complete student survey instruments, including response options and item-level summaries/visualizations; and  (S3) the complete faculty survey questions with item-level response summaries and visualizations. 

Table \ref{tab:survey_comparison} provides a high-level overview of the thematic structure of the student and faculty survey instruments, highlighting the parallel but role-specific design of the two surveys. 
Table \ref{tab:survey_mapping} complements this overview by providing a mapping of each survey item, whether its primary results are presented in the main manuscript or in the supplementary materials. Together, these tables are intended to improve transparency, clarify the organization of the instruments, and help readers efficiently locate the full wording of questions, response options, and associated results.

\begin{table}[H]
\footnotesize
\centering
\caption{Students and faculty survey structure.}
\label{tab:survey_comparison}
\begin{tabular}{m{4.5cm} c m{6cm} c}
\toprule
\multicolumn{2}{c}{\textbf{Student Survey}} & \multicolumn{2}{c}{\textbf{Faculty Survey}} \\
\midrule
Section & \# Questions & Section & \# Questions \\
\midrule
Consent & 2 & Consent & 2 \\
\addlinespace
Demographics \& Background & 7 & Background & 6 \\
\addlinespace
Awareness of AI Tools for DS & 5 & Competency in Using AI Tools for Teaching DS & 2 \\
\addlinespace
Perceptions Toward AI in DS & 3 & Perceptions Toward AI in DS Education & 2 \\
\addlinespace
Understanding AI Limitations and Concerns & 3 & Challenges and Concerns Regarding AI in DS Education & 2 \\
\addlinespace
Potential to Improve the DS Workflow with AI Tools & 4 & Potential of AI to Improve DS Teaching & 1 \\
\addlinespace
& & Ethical Considerations and Institutional Policies & 4 \\
\addlinespace
& & AI Readiness and Faculty Support Needs & 3 \\
\bottomrule
\end{tabular}
\end{table}

\begin{table}[H]
\centering
\caption{Mapping of student and faculty survey items to the main manuscript and supplementary materials.}
\label{tab:survey_mapping}
\scriptsize
\renewcommand{\arraystretch}{1.18}
\setlength{\tabcolsep}{4pt}

\begin{tabularx}{\textwidth}{p{0.13\textwidth} X|p{0.13\textwidth} X}
\toprule
\multicolumn{2}{c|}{\textbf{Student Survey}} & \multicolumn{2}{c}{\textbf{Faculty Survey}} \\
\midrule
\textbf{Question Number} & \textbf{Location / Notes} & \textbf{Question Number} & \textbf{Location / Notes} \\
\midrule

Q1--Q2 & Administrative consent items. 
& Q1--Q2 & Administrative consent items. \\

Q3--Q5 & Demographic and background items; full wording and response options provided in the supplement. 
& Q3--Q6 & Background items; full wording and response options provided in the supplement. \\

Q6--Q7 & Primary results discussed in the main text; full question wording and corresponding supplementary figures also provided in the supplement. 
& Q7--Q8 & Primary results discussed in the main text; supplementary figures and full question wording provided in the supplement. \\

Q8--Q9 & Presented in the supplement with full wording and response summaries. 
& Q9--Q11 & Primary results are presented in the main text; Q11 figure is in the supplement. Full question wording is provided in the supplement. \\

Q10--Q14 & Primary results are discussed in the main text; detailed response distributions and full question wording are also provided in the supplement. 
& Q12 & Primary result discussed in the main text; supplementary figure and full question wording provided in the supplement. \\

Q15--Q17 & Q16--Q17 are discussed in the main text; the rest are presented in the supplement with full wording and response summaries. 
& Q13--Q15 & Primary results are discussed in the main text; for Q13 and Q15, figures are not duplicated in the supplement. Full question wording is provided in the supplement. \\

Q18--Q19 & Primary results are discussed in the main text; full question wording and supplementary response summaries are provided in the supplement. 
& Q16 & Omitted during survey validation due to redundancy. \\

Q20 & Presented in the main text and supplement with full wording and response summary. 
& Q17--Q23 & Primary results are discussed in the main text; supplementary figures and full question wording are provided in the supplement, except that figures for items with primary main-text figures are not duplicated. \\

Q21 & Primary result discussed in the main text; full question wording provided in the supplement. 
&  &  \\

Q22--Q23 & Discussed in both the main text and the supplement with full wording and response summaries. 
&  &  \\

Q24 & Primary result discussed in the main text; full question wording and supplementary figure provided in the supplement. 
&  &  \\
\bottomrule
\end{tabularx}

\vspace{0.5em}
\parbox{0.97\textwidth}{\scriptsize \textit{Note.} Questions Q1--Q2 in both instruments were administrative consent items. Survey questions were shown only to participants who met the admission requirements and consented to participate. For the faculty survey, figures for items whose primary visual results appear in the main text (Q9, Q10, Q13, and Q15) are not duplicated in the supplement. Faculty survey item Q16 was removed during survey validation due to redundancy. }
\end{table}

\section{Supplementary Results}
\title{Supplementary Results and 
Exploratory Analysis}

This section presents supplementary analyses that support the primary findings reported in the main text. The results are organized to parallel Sections~3 and~4 of the main paper.

\subsection{Detailed Student Familiarity with AI Tools (Supplementary to Section~3.1)}

{To complement the summary of student familiarity reported in Section~3.1 of the main text, we present more granular results on students' familiarity with specific AI tools and task-level AI use.}

{\textbf{Tool-specific familiarity.} Students rated their familiarity with five AI technologies on a scale from ``Not at All Familiar'' to ``Very Familiar'' (Q10). As reported in the main text, familiarity was highest for general-purpose tools and lowest for domain-specific platforms. Figure~\ref{fig:familiarity_likert} (Section~S2 of this supplement) shows the full distribution. ChatGPT had the highest proportion of students reporting familiarity (57\% ``Somewhat Familiar'' or above), while AI-driven analytics platforms had the lowest, with 56\% of students reporting no familiarity at all. Gemini and Google Colab occupied intermediate positions, with roughly one-third of students reporting at least some familiarity. These patterns suggest that students' exposure to AI is shaped primarily by the accessibility and general visibility of tools rather than by disciplinary relevance.}

{\textbf{Tools used in DS tasks.} When asked which AI tools they had used in their data science tasks (Q11), ChatGPT again dominated (80\%), followed by Grammarly (53.6\%) and AI-powered visualization tools (26.4\%). Only 4.5\% reported using predictive analytics tools such as DataRobot, and 10\% reported using no AI tools at all. Figure~\ref{fig:aQ11_ai_tools_used} (Section~S2) shows the full distribution. The low adoption of specialized analytics platforms, despite their direct relevance to DS workflows, further underscores the gap between general-purpose and domain-specific AI literacy identified in the main text.}

{\textbf{Use cases for AI.} Students reported using AI technology primarily for coding assistance (69.4\%), proofreading (52.3\%), mathematical proofs (36.9\%), literature reviews (30.6\%), and generating initial drafts (27.9\%). Figure~\ref{fig:Q7} (Section~S2) presents these results. Please note that students were permitted to provide multiple responses to this question; the relative frequency is calculated using the number of nonempty responses ($n=111$) as the denominator. The predominance of implementation-oriented use cases over content-generation tasks supports the main text's finding that students engage with AI primarily for refinement and support rather than for producing original analytical content.}

\subsection{Detailed Student Perceptions of AI in DS (Supplementary to Section~3.2)}

{Section~3.2 of the main text reports that students are broadly optimistic about AI's potential in data science. This subsection provides additional detail on perceptions toward specific AI capabilities, future career relevance, and the preferred role of AI relative to traditional methods.}

{\textbf{Agreement with AI capability statements.} Students rated their agreement with four statements about AI tools in DS (Q12). Figure~\ref{fig:Q12_attitudes} (Section~S2) shows the response distributions. Agreement was strongest for the statements that AI tools can automate tedious data processing (71\% Agree or Strongly Agree) and provide new insights (69\%). Regarding the statement that AI tools lack emotional intelligence, responses were more divided: 55\% agreed, while 36\% were neutral, reflecting recognition of a limitation alongside the generally positive outlook.}

{\textbf{Perceptions of the Future of AI in DS.} Students also rated their agreement with statements about AI's future role in DS careers (Q15). Figure~\ref{fig:Q15_attitudes} (Section~S2) presents the full distributions. More than 65\% agreed that AI tools are essential for future DS workflows and that AI improves efficiency in analyzing large datasets. Agreement was also high for the statement that AI makes DS accessible to non-technical professionals (75\% Agree or Strongly Agree). However, the statement that over-reliance on AI could diminish fundamental DS skills received the highest proportion of strong agreement (36\% Strongly Agree), indicating that optimism about AI's potential coexists with awareness of its risks.}

{\textbf{Willingness to learn and adopt AI tools.} When asked whether students should learn AI tools for DS careers (Q16), 57.8\% selected ``Yes, but not a priority'' and 32.4\% selected ``Yes, it is essential.'' No students selected ``No, it is not necessary'' or ``No, it may even be harmful.'' Regarding willingness to incorporate AI tools into future DS projects (Q17), 72.4\% agreed or strongly agreed, while only 7.7\% disagreed or strongly disagreed.}

{\textbf{Complement vs.\ replace traditional tools.} Figure~\ref{fig:Q24} (Section~S2) shows how students responded when asked whether AI tools should complement or replace traditional DS methods (Q24). A majority (54.9\%) endorsed maintaining a balance between AI tools and traditional approaches, and 31.4\% favored a primarily complementary role. Only 5.9\% indicated that AI mostly replaces traditional methods, and 2.0\% preferred full reliance on AI. An equal proportion (5.9\%) reported that AI tools do not meaningfully complement traditional data science tools. These results reinforce the main-text finding that students favor integrative rather than substitutive adoption of AI tools.}

\subsection{Detailed Student Awareness of AI Limitations (Supplementary to Section~3.3)}

{The main text reports that at least 70\% of students recognize key AI limitations and that accuracy and reliability are the most frequently cited concerns. This subsection provides the detailed response distributions for awareness items, concern rankings, and integration challenges.}

{\textbf{Encounters with AI issues.} When asked whether they had encountered hallucinations, bias, or inaccuracy while using AI tools for DS tasks (Q20), 53.4\% reported occasional encounters and 15.5\% reported frequent encounters. Only 18.4\% reported never encountering such issues, while 12.6\% had never used AI tools for DS.}

{\textbf{Primary concerns about AI in DS.} Students were asked to identify their primary concerns about using AI tools in DS (Q19; multiple responses allowed). Figure~\ref{fig:Q19_Concerns} (Section~S2) shows the distribution. Accuracy and reliability were cited most frequently (80.6\%), followed by privacy and security (63.1\%), skill degradation (62.1\%), ethical concerns or bias (58.3\%), lack of transparency (39.8\%), and difficulty understanding the models (34\%). These responses suggest that accountability and transparency are salient concerns for students.}

{\textbf{Challenges in integrating AI tools into DS workflows.} Students were also asked about specific challenges they face when integrating AI tools into their work (Q22). Figure~\ref{fig:Q22-challenges} (Section~S2) shows the distribution. Accuracy and reliability of predictions were the most frequently cited challenges (48\%), followed by data security and privacy risks (19.6\%), understanding AI outputs (16.7\%), and complexity of using AI tools (8.8\%). Only 6.9\% expressed concern about AI's impact on DS job roles. These results suggest that students' primary integration challenges are epistemic (can I trust the output?) rather than occupational (will AI replace my job?).}

{\textbf{Concerns about AI in education.} Figure~\ref{fig:Studentconcerns} (Section~S2) presents students' concerns about integrating AI into education more broadly (Q23). Over-reliance on generative AI was the most commonly selected concern (42.2\%), followed by hindering the development of transferable skills (23.5\%), undermining the value of university education (18.6\%), and limiting social interaction (15.7\%). The prominence of over-reliance as a concern, echoed in the faculty results, suggests a shared awareness across stakeholder groups that AI adoption requires deliberate safeguards against dependency.}\\


\subsection{Detailed Faculty Integration and Self-Assessment (Supplementary to Section~4.1)}

{Section~4.1 of the main text reports that a majority of faculty have incorporated AI tools into teaching at least sometimes, but that self-assessed proficiency remains low. This subsection provides the detailed distributions for specific teaching tasks and usage frequency.}

{\textbf{Faculty AI incorporation.} When asked whether they had incorporated AI tools in their data science teaching (Q7), 57.1\% of faculty responded ``Sometimes,'' 14.3\% ``Rarely,'' and 7.1\% ``Always.'' Only 21.4\% had never incorporated AI tools. Figure~\ref{fig:Q7_used_AI} (Section~S3 of this supplement) shows the distribution.}

{\textbf{AI tools and platforms used.} When asked which AI tools or platforms they use in teaching (Q8), AI learning assistants such as ChatGPT dominated the list (100\% of respondents), followed by AI-driven coding platforms such as Jupyter and Colab (64.3\%) and AI-based data analysis or visualization tools (42.9\%). Only 14.3\% reported using automated grading systems, and one respondent mentioned AI project management or collaboration tools. These findings are consistent with student responses, in which ChatGPT was the most popular GenAI tool.}

{\textbf{Self-rated ability for specific teaching tasks.} Faculty rated their ability to perform five AI-related teaching tasks (Q9). Across all five tasks, the majority rated their ability as ``Beginner'' or ``No Experience.'' The task with the highest reported competence was implementing AI-assisted coding environments, with 40\% reported at least intermediate skill. The lowest self-rated competence was for employing AI for student performance tracking and curriculum adjustments, with 62\% reporting no experience. Figure~4 in the main text summarizes these results. The gap between faculty self-assessment and student self-assessment is noteworthy. Whether this reflects greater realism among faculty or greater overconfidence among students cannot be determined from survey data alone, but it has implications for calibrating training programs and managing expectations in AI-integrated courses.}

{\textbf{Frequency of AI use in specific teaching activities.} Faculty rated how frequently they use AI for four teaching activities (Q10). Across all activities, ``Never'' or ``Rarely'' was the most common response, selected by over 46\% of faculty for each item. ``Generating teaching materials with AI support'' and ``Engaging students through AI-assisted interactive labs'' had the highest frequency of use, although even for these tasks fewer than 23\% of faculty reported using AI ``Always'' or ``Often.'' Figure~5 in the main text presents the full distributions.}

\subsection{Detailed Faculty Perceptions (Supplementary to Section~4.2)}

{The main text highlights that faculty perceptions are broadly supportive but conditional. This subsection provides additional detail on the equity and inclusion finding and on perceptions of AI as an instructional enhancer.}

{\textbf{Equity and inclusion.} A notable finding was that faculty were unanimous in the direction of their response to the statement ``AI tools are capable of improving equity and inclusion in data science education'': no faculty disagreed, 23\% were neutral, and the remainder agreed or strongly agreed. This result is particularly striking given widespread attention to algorithmic bias and the risk that GenAI tools may amplify existing societal and racial inequities. As an HBCU, our institutional context makes this finding especially relevant. Faculty responses were favorable toward the potential of GenAI tools to improve equity and inclusion in DS education, even as they recognized the risks of bias. This finding should be interpreted cautiously, given the small sample size. }

{\textbf{Detailed Perceptions toward AI in DS Education.} Figure~6 in the main text summarizes faculty perceptions across five pedagogical and ethical dimensions (Q11). A strong majority expressed agreement that AI tools can improve equity and inclusion (over 75\%), efficiently automate repetitive teaching tasks (nearly 70\%), and significantly improve instruction in technically complex areas such as machine learning, deep learning, and big data (over 75\%). Likewise, a majority agreed that AI tools will be vital for the future of DS education. Only 9\% disagreed with the claim that GenAI tools can efficiently automate repetitive teaching tasks. The item on over-reliance reducing critical thinking skills was more evenly distributed across agreement and neutrality, indicating ambivalence rather than outright opposition.}

{\textbf{AI tools as teaching enhancers.} Figure~7 in the main text summarizes faculty agreement that AI tools can enhance five dimensions of DS teaching (Q15). The strongest endorsement was for simplifying complex DS concepts through AI-enhanced visualizations (roughly two-thirds agreeing or strongly agreeing). No faculty disagreed with the potential for AI to support real-time tracking and identification of student challenges, indicating strong confidence in AI's role in formative assessment and conceptual support. Faculty also viewed AI favorably for enhancing engagement via interactive modules and automating assessment and feedback, although these areas exhibited slightly higher neutrality and some disagreement, suggesting cautious optimism, particularly regarding automated assessment, where concerns about validity or pedagogical fit may remain. Perceptions were most mixed regarding personalizing learning paths using AI-driven insights, with the highest proportion of neutral and negative responses.}

{\textbf{Complement vs.\ replace (faculty).} When asked whether AI tools should complement or replace traditional methods in teaching DS (Q12), 61.5\% of faculty favored balancing AI tools with traditional methods, 30.8\% preferred AI as a complement, and only 7.7\% supported replacement in some contexts. No faculty supported full replacement, a pattern consistent with the student results. Figure~\ref{fig:Q12_Complement_vs_Replace} (Section~S3 of this supplement) shows the distribution.}

\subsection{Detailed Faculty Concerns and Institutional Context (Supplementary to Section~4.3)}

{The main text summarizes the headline faculty concerns and syllabus policy findings. This subsection provides additional detail on the distribution of concern levels across categories, institutional awareness, ethical priorities, and training needs.}

{\textbf{Concern levels by category.} Figure~8 in the main text presents the full distribution of faculty concern across five categories (Q13). Beyond the headline finding that all faculty were concerned about keeping pace with evolving AI technologies, the data reveal that concerns about biased results in student assessments and lack of transparency in AI outputs were nearly as prevalent, with over 70\% expressing concern in each area. Over-reliance on AI reducing faculty-student engagement was also a majority concern, with 23\% not concerned and 46\% somewhat concerned. Data privacy and security, while still a majority concern, had the lowest intensity, with 8\% of faculty reporting no concern. This relatively lower concern about privacy may reflect greater familiarity with data-driven technologies among DS faculty.}

{\textbf{Integration challenges.} When asked about their biggest challenges in integrating AI tools into their courses (Q14), faculty most frequently cited insufficient institutional support or training (61.5\%), balancing AI with curriculum learning objectives (53.8\%), ethical concerns such as fairness and bias (46.2\%), resistance from students or faculty to adopt AI tools (38.5\%), limited access to advanced AI platforms (38.5\%), and difficulty in learning and mastering AI tools (23.1\%). Figure~\ref{fig:Q14_Challenges_AI} (Section~S3 of this supplement) presents the full distribution.}

{\textbf{Institutional AI guidelines awareness.} Figure~\ref{fig:Q17_faculty_awareness_AI_guidelines} (Section~S3) shows that only 7.7\% of faculty reported being ``Very Aware'' of institutional guidelines on AI use. The largest group (46.2\%) reported being ``Somewhat Aware,'' while 38.5\% reported being ``Not Aware.'' This limited awareness, combined with the non-uniform syllabus policies reported in the main text, suggests a significant gap in institutional communication regarding AI governance in the classroom.}

{\textbf{Ethical priorities.} Figure~\ref{fig:Q18_Primary_Ethical_Concern} (Section~S3) shows that 84.6\% of faculty cited over-reliance on AI leading to less human oversight as their primary ethical concern, while 15.4\% cited bias in AI-driven student evaluations. No faculty member selected student privacy as their primary concern.}

{\textbf{Human oversight.} Faculty unanimously endorsed the importance of maintaining human oversight when using AI for assessments and student feedback (Q19). Figure~\ref{fig:Q19_Human_Oversight} (Section~S3) shows that 76.9\% rated this as ``Very Important'' and 23.1\% as ``Important.''}

{\textbf{Syllabus policies.} Figure~\ref{fig:Q20_AI_Policy_Syllabus} (Section~S3) presents the distribution of AI policies in faculty syllabi (Q20). The majority (83.3\%) allowed AI use for specific activities designated by the instructor with required disclosure, 16.7\% prohibited AI use for any activities or assignments, and none permitted unrestricted use with attribution.}

{\textbf{Faculty confidence.} When asked about their confidence in their knowledge of AI tools available for DS education (Q21), responses were distributed across all levels: 30.8\% ``Very Confident,'' 15.4\% ``Somewhat Confident,'' 30.8\% ``Neutral,'' and 23.1\% ``Not Confident.'' Figure~\ref{fig:Q21_confidence_in_AI_tools} (Section~S3 of this supplement) presents these results.}

{\textbf{Training and support needs.} Figure~\ref{fig:Fac-training-Needs_Q22} (Section~S3) presents the distribution of faculty training needs (Q22). Access to advanced AI platforms for classroom use and technical support for integrating AI into the curriculum were each cited by 69.2\% of faculty. More AI-related training or professional development was selected by 61.5\%, institutional support for purchasing AI tools by 53.8\%, and peer collaboration and sharing best practices by 46.2\%.}

{\textbf{Institutional support adequacy.} When asked whether their institution provides sufficient guidance and resources for using AI in teaching (Q23), 53.8\% responded ``No,'' 30.8\% responded ``Somewhat,'' and only 15.4\% responded ``Yes.'' Figure~\ref{fig:Q23_Institutional_Guidance} (Section~S3) presents these results. This finding reinforces the main text's conclusion that institutional leadership and support for AI integration are significant areas for improvement.}

\section{Student Survey Questions}\label{sec:student_survey}

\title{Student Survey Questions}

 This section includes detailed student survey questions, response options, and additional secondary results or visualizations.

\newlist{qenum}{description}{1}
\setlist[qenum]{
  style=nextline,
  font=\normalfont\bfseries,
  leftmargin=2.4em,
  labelsep=0.6em,
  itemsep=1.15em,
  topsep=0.5em
}

\setlist[itemize]{
  leftmargin=1.8em,
  itemsep=0.25em,
  topsep=0.25em
}

\newcommand{\resp}[1]{\par\smallskip\noindent\hspace*{2.4em}\textit{Response options:} #1}
\newcommand{\commentspace}{\vspace{1.0em}}

\newcommand{\LikertSA}{\emph{(Strongly disagree--Strongly agree)}}
\newcommand{\LikertFam}{\emph{(Not at all familiar--Very familiar)}}
\newcommand{\LikertFreq}{\emph{(Never--Always)}}
\newcommand{\LikertAware}{\emph{(Not at all aware--Very aware)}}

\vspace{1em}

\begin{qenum}

\item[Q3.] What is your gender?
\resp{Male; Female; Non-binary; Trans man; Trans woman; Prefer not to say.}
\commentspace

\item[Q4.] What is your major?
\resp{Open-ended response.}
\commentspace

\item[Q5.] What is your classification?
\resp{Freshman; Sophomore; Junior; Senior; Graduate student.}
\commentspace

\item[Q6.] Have you ever used AI technologies (e.g., MS Copilot, Google Colab) in your data science work?
\resp{Never; Rarely; Sometimes; Often; Very often.}
\commentspace
 \begin{figure}[H]
    \centering
     \includegraphics[width=0.8\textwidth]{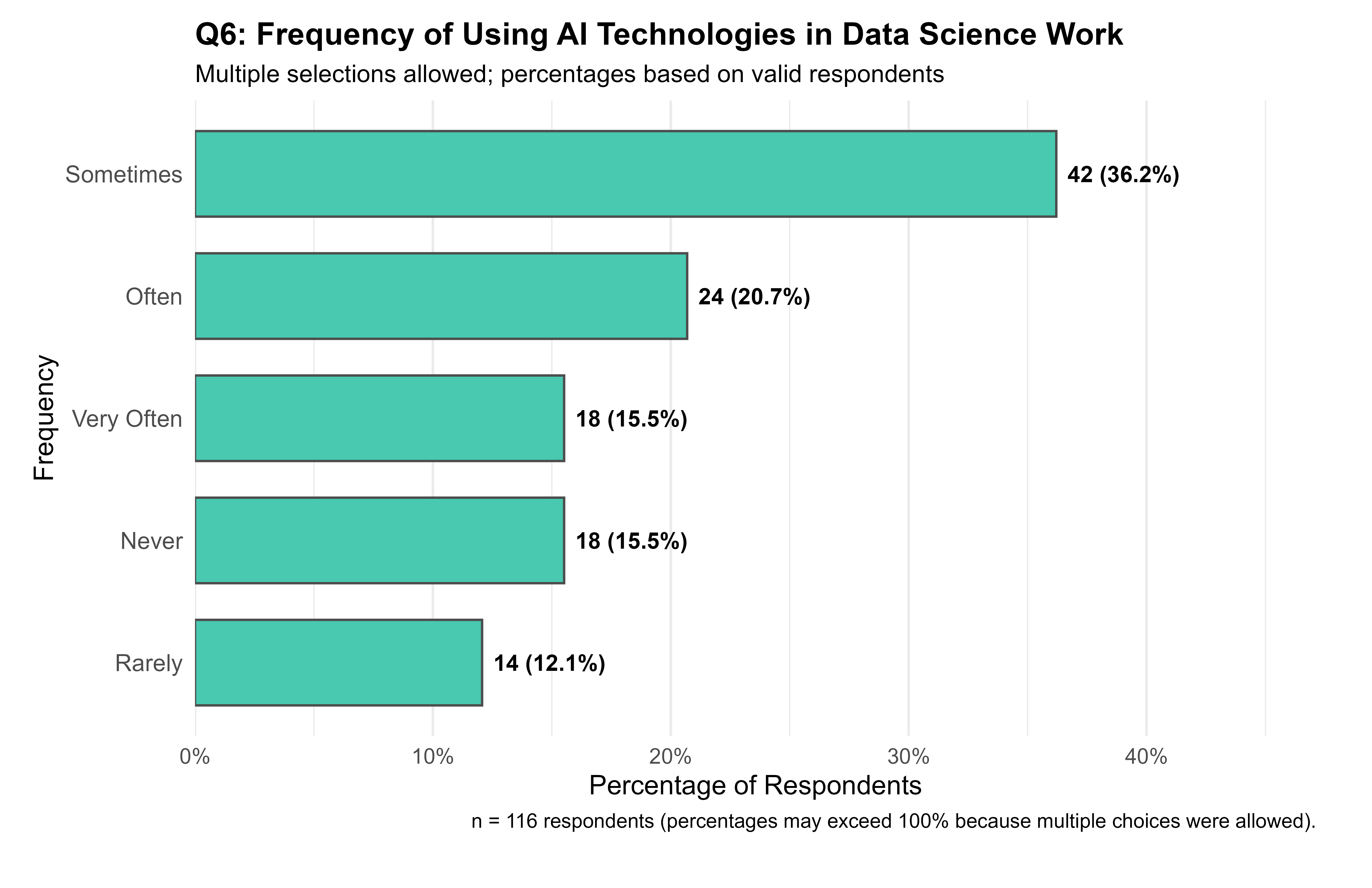}
        \caption{Students' self-reported frequency of using AI tools in their DS coursework.} 
        \label{fig:Q6}
\end{figure}

\item[Q7.] What did you use AI technology for? Select all that apply.
\resp{Proofreading; Getting an initial draft; Help with coding; Mathematical proofs; Literature review; Other (please specify).}
\commentspace
\begin{figure}[H]
    \centering
    \includegraphics[width=0.8\textwidth]{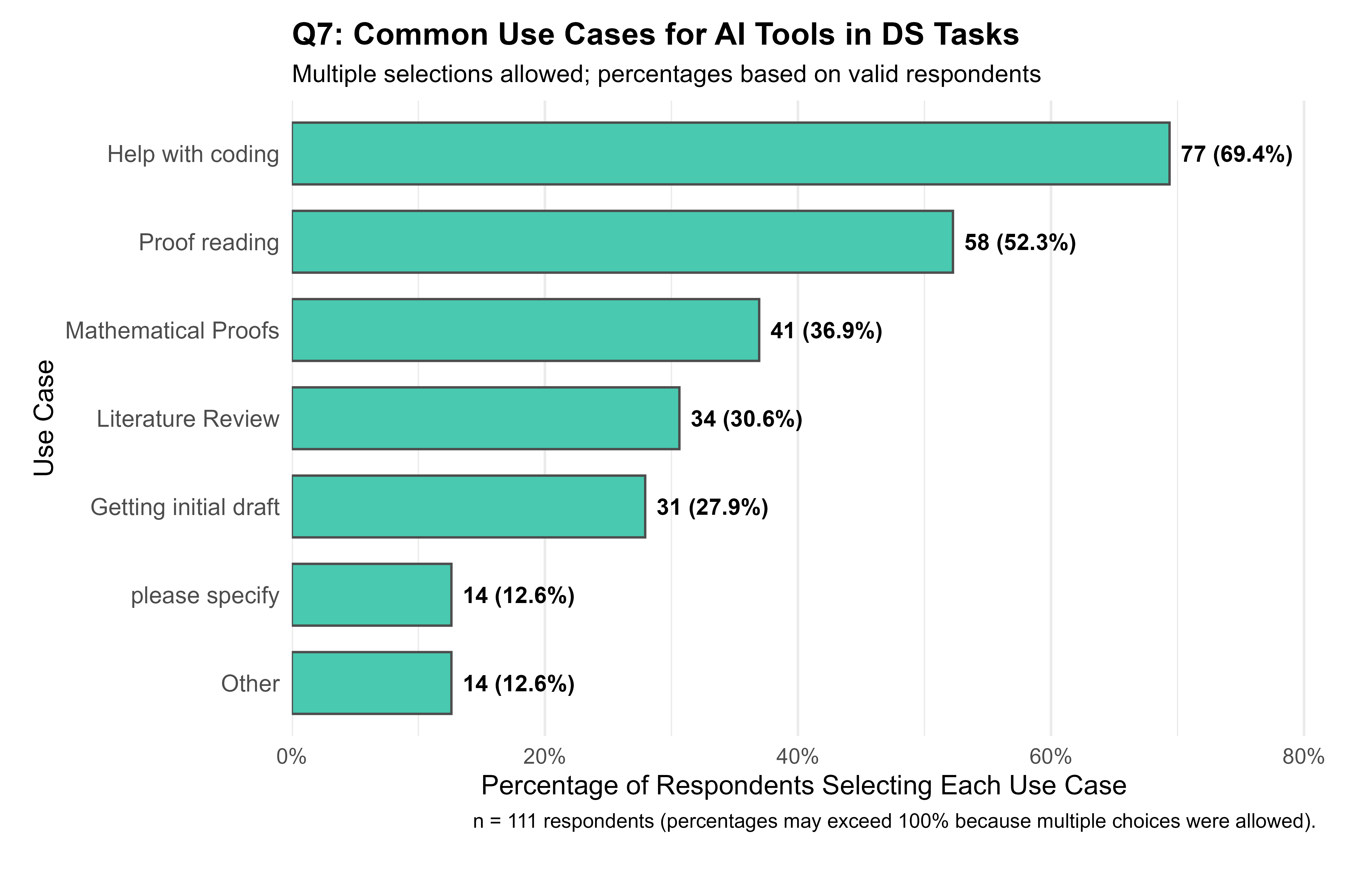}
    \caption{Students' common use cases for AI tools in DS tasks.}
    \label{fig:Q7}
\end{figure}

\item[Q8.] Are you interested in pursuing a degree in data science in the future?
\resp{Yes; Not sure; No.}\\
\commentspace
Among respondents, (40.9\%) selected ``Yes,'' 36.4\% selected ``No,'' and 22.7\% were unsure.

\item[Q9.] Are you interested in getting a job in data science in the future?
\resp{Yes; Not sure; No.}\\
\commentspace
Among respondents, (46.8\%) selected ``Yes,'' 27.9\% selected ``No,'' and 25.2\% were unsure.

\item[Q10.] How familiar are you with the following AI technologies in data science?
\resp \LikertFam
\begin{itemize}
    \item ChatGPT 
    \item Gemini 
    \item Google Colab 
    \item AI-powered data visualization tools (e.g., Tableau, Power BI with AI) 
    \item AI-driven analytics platforms (e.g., DataRobot) 
\end{itemize}
\commentspace

\begin{figure}[H]
    \centering
    \includegraphics[width=0.95\textwidth]{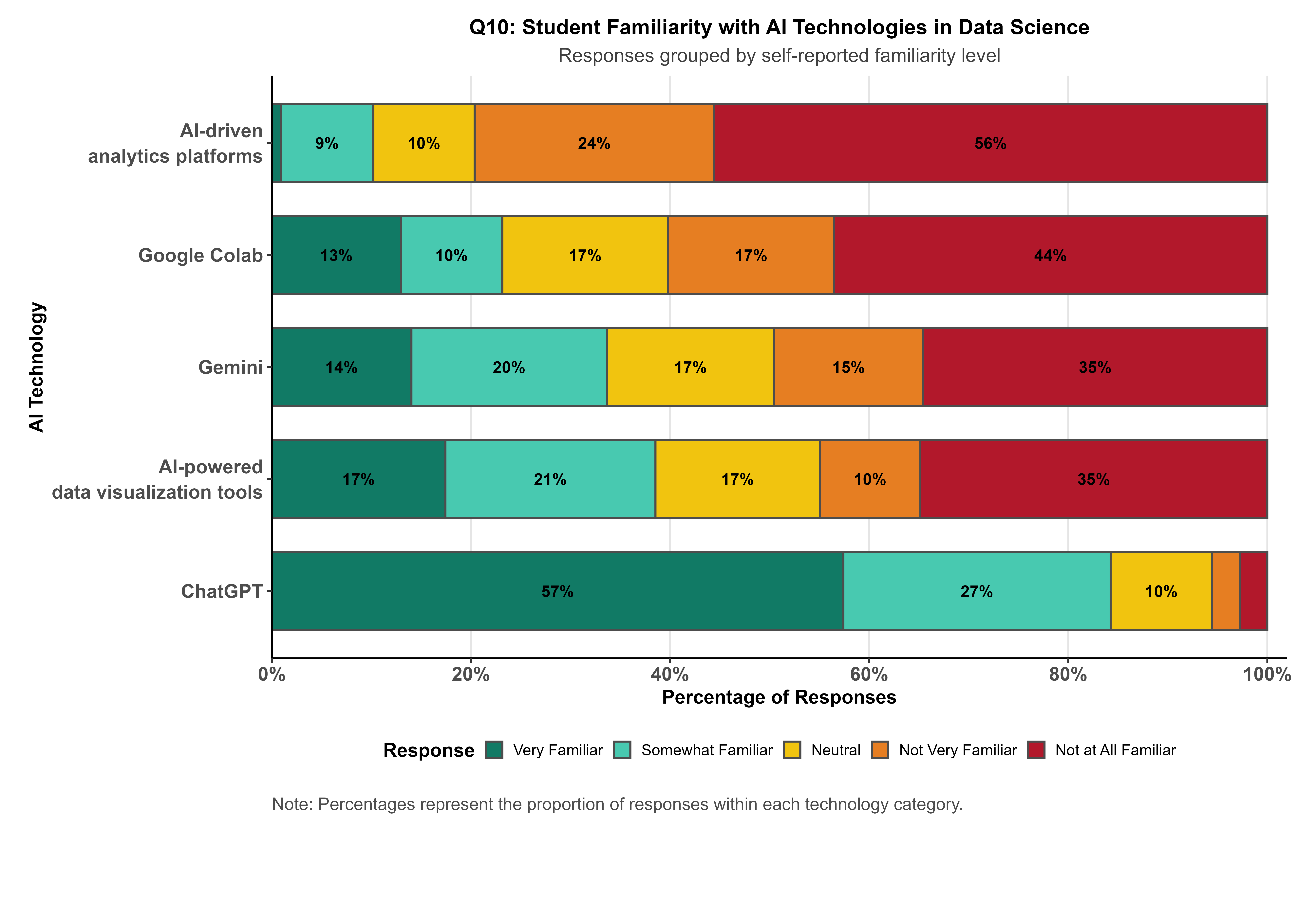}
        \caption{Students' self-reported familiarity levels with AI technologies in DS.}
    \label{fig:familiarity_likert}
\end{figure}


\item[Q11.] Which AI tools have you used in your data science tasks? Select all that apply.
\resp{ChatGPT; AI-powered visualization tools; Predictive analytics tools (e.g., DataRobot); Grammarly; Canva or AI-assisted presentation tools; Other (please specify); None.}
\commentspace
\begin{figure}[H]
    \centering
    \includegraphics[width=0.85\textwidth]{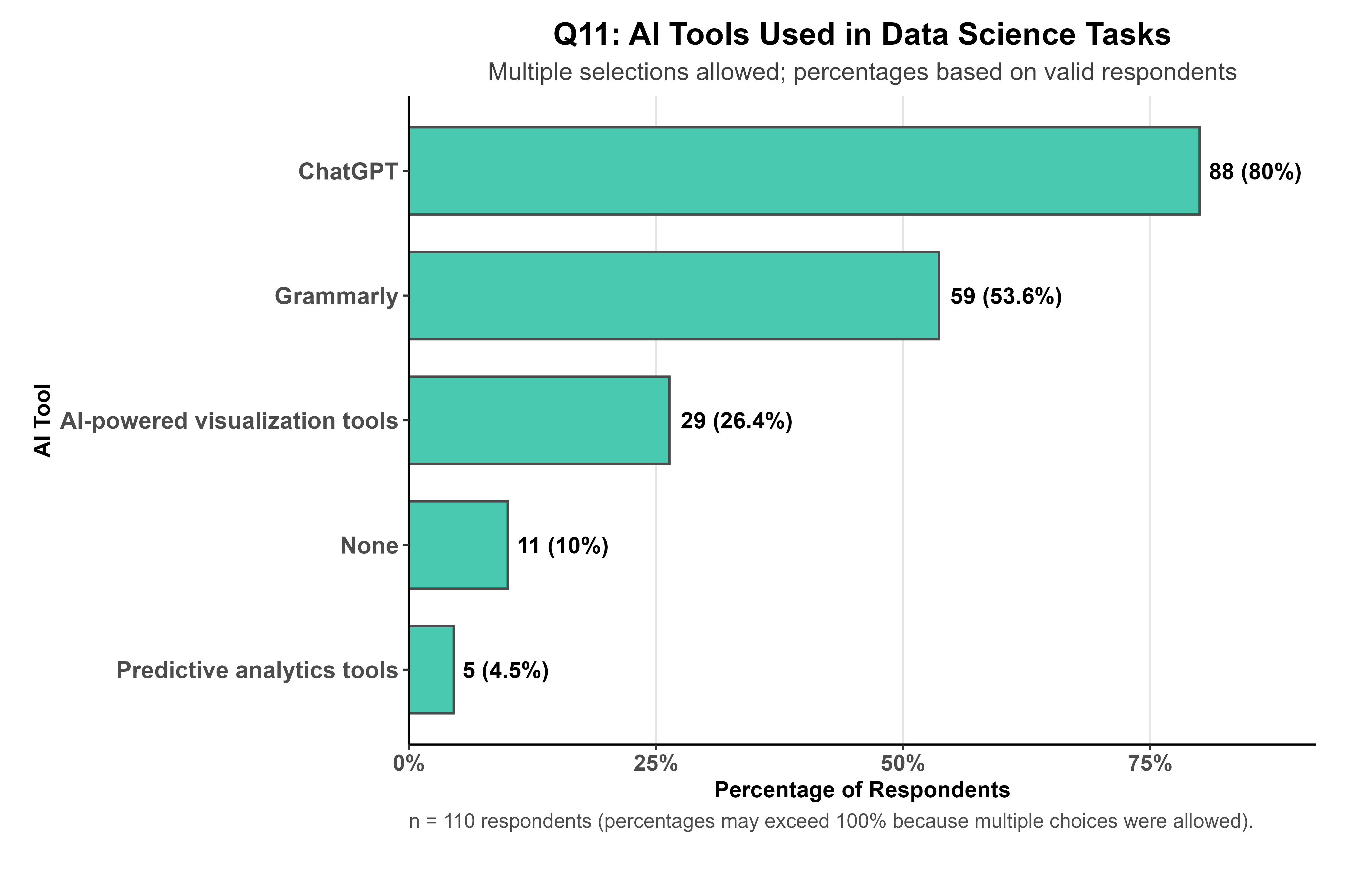}
        \caption{AI tools students reported using in DS tasks.}
    \label{fig:aQ11_ai_tools_used}
\end{figure}

\item[Q12.] To what extent do you agree with the following statements regarding AI tools in data science?
\resp \LikertSA
\begin{itemize}
    \item AI tools can enhance the accuracy of data science models 
    \item AI tools can automate tedious data processing tasks 
    \item AI tools can provide new insights or perspectives I would not have discovered otherwise 
    \item AI tools lack emotional intelligence, which may limit their utility in complex tasks 
\end{itemize}
\commentspace
\begin{figure}[H]
    \centering
    \includegraphics[width =.95\textwidth]{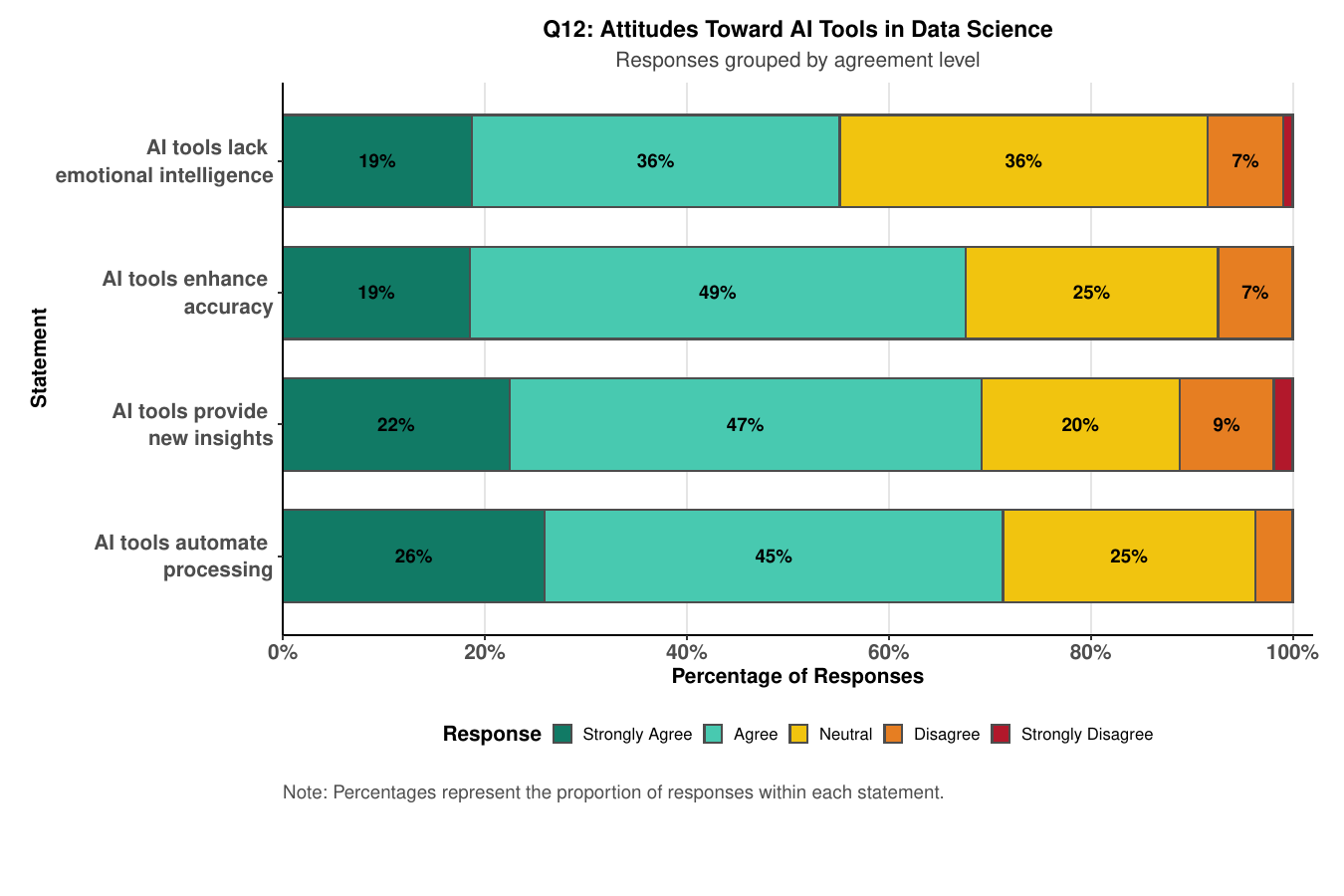}
    \caption{Students' perceptions toward integrating AI tools in DS.}
    \label{fig:Q12_attitudes}
\end{figure}

\item[Q13.] How often do you use AI tools for the following data science tasks?
\resp \LikertFreq
\begin{itemize}
    \item Exploratory data analysis 
    \item Predictive modeling 
    \item Data visualization 
    \item Automating data cleaning tasks 
    \item Presenting findings and reports 
\end{itemize}
\commentspace

\item[Q14.] How confident are you in explaining AI-generated outputs for data analysis tasks to peers or supervisors?
\resp{Not at all confident; Not very confident; Neutral; Confident; Very confident.}

\commentspace
Only 40.6\% of student respondents were confident in their ability to explain AI outputs, while 30.2\% were either not very confident or not at all confident. 

\item[Q15.] To what extent do you agree with the following statements regarding the future of AI in data science?
\resp \LikertSA
\begin{itemize}
    \item AI tools are essential for future data science workflows 
    \item AI can improve efficiency in analyzing large datasets 
    \item AI can make data science more accessible to non-technical professionals 
    \item Overreliance on AI tools could diminish fundamental data science skills 
\end{itemize}

\commentspace
\begin{figure}[H]
    \centering
    \includegraphics[width=0.8\textwidth]{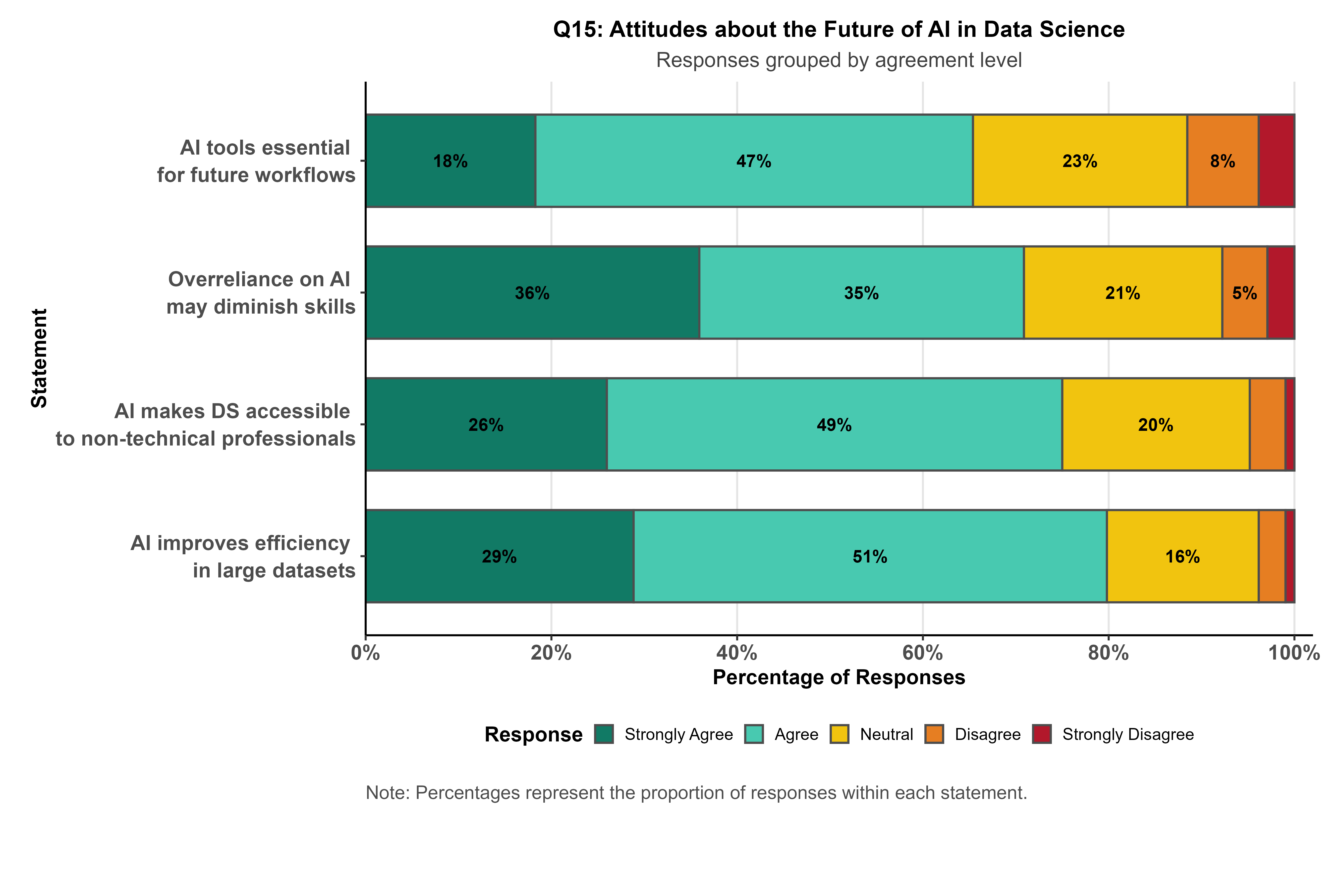}
    \caption{Agreement with statements about the future of GenAI in  DS.}
    \label{fig:Q15_attitudes}
\end{figure}

\item[Q16.] Do you think students should learn how to use AI tools for their data science careers?
\resp{Yes, it is essential; Yes, but not a priority; Neutral; No, it is not necessary; No, it may even be harmful.}

\commentspace

The majority of students (57.8\%) selected ``Yes, but not a priority'', about a third (32.4\%) selected ``Yes, it is essential''. The rest selected ``Neutral''. Neither the option ``No, it is not necessary'' nor the option ``No, it may even be harmful'' was selected by any students.

\item[Q17.] I am willing to incorporate AI tools more into my data science projects in the future.
\resp{Strongly disagree; Disagree; Neutral; Agree; Strongly agree.}\\

\commentspace
The majority of students indicated willingness to incorporate AI tools into their future DS projects, with 20\% selecting ``Strongly Agree,'' 52.4\% selecting ``Agree,'' 6.7\% selecting ``Disagree,'' and 1\% selecting ``Strongly Disagree.'' 

\item[Q18.] To what extent are you aware of the following limitations of using AI for data science tasks?
\resp \LikertAware
\begin{itemize}
    \item AI tools can produce biased results due to biased training data 
    \item AI tools may generate factually incorrect outputs (hallucinations)  
    \item AI models lack real-world experience and contextual understanding  
    \item AI-generated results often need to be validated by experts  
    \item Using AI tools could pose data privacy risks 
\end{itemize}
\commentspace

\item[Q19.] What are your primary concerns about using AI tools in data science? Select all that apply.
\resp{Accuracy and reliability of AI outputs; Lack of transparency in AI decision-making; Data privacy and security issues; Ethical concerns (e.g., bias in AI models); Overreliance on AI leading to skill degradation; Difficulty understanding how AI models work; Don’t know/none.}
\commentspace
\begin{figure}[H]
    \centering
    \includegraphics[width=0.85\textwidth]{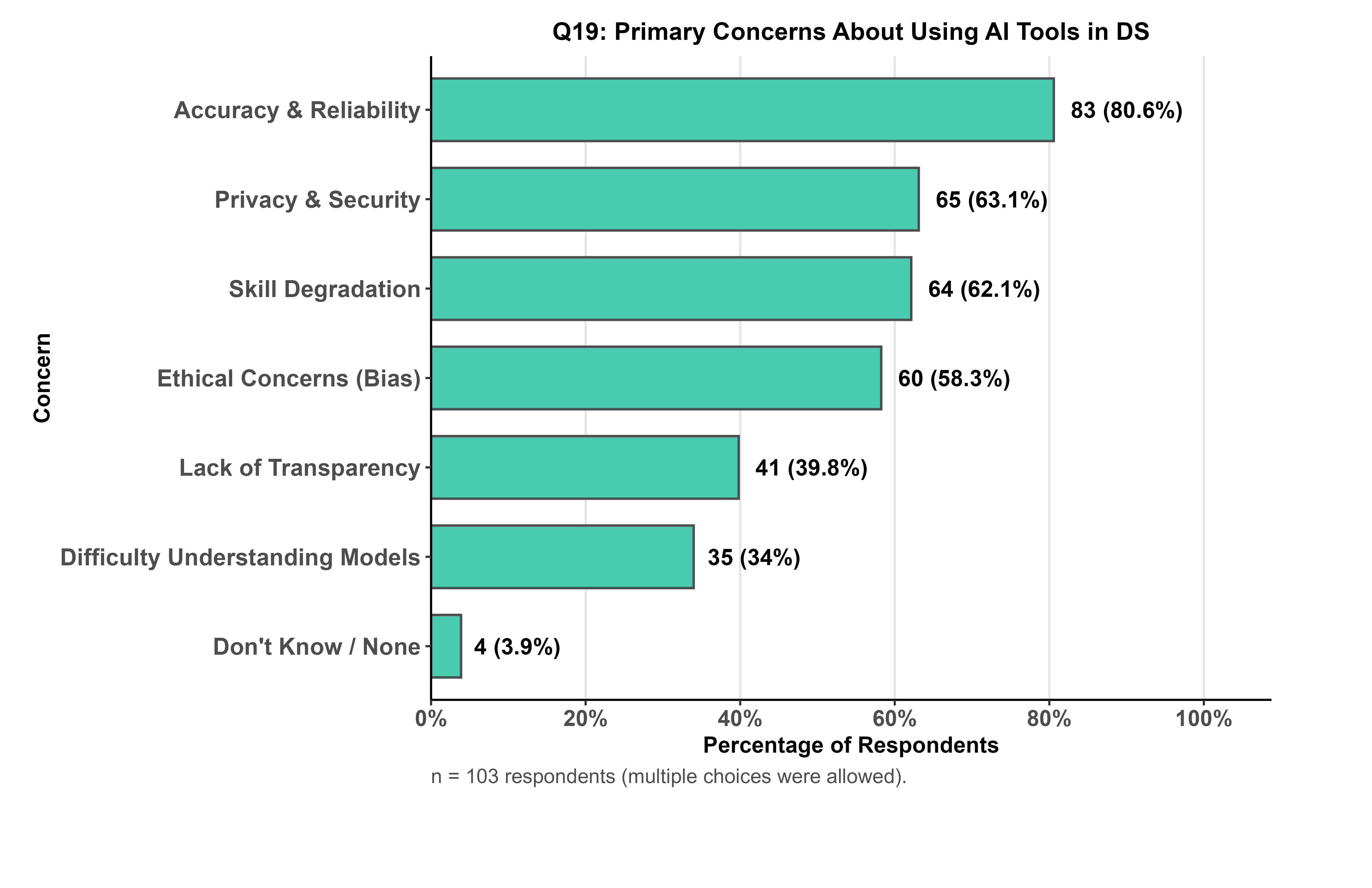}
    \caption{Students' primary concerns about using AI tools in DS.}
    \label{fig:Q19_Concerns}
\end{figure}

\item[Q20.] Have you ever encountered issues such as hallucinations, bias, or inaccuracy while using AI tools for data science tasks?
\resp{Yes, frequently; Yes, occasionally; No, never; I have never used AI tools for data science.}\\
\commentspace
The majority (53.4\%) selected ``Yes, occasionally'', 15.5\% selected ``Yes, frequently'', 18.4\% selected ``No, never'', and 12.6\% selected ``I never used AI tools for data science.''

\item[Q21.] Do you think AI tools can significantly enhance the following aspects of your data science workflow?
\resp \LikertSA
\begin{itemize}
    \item Automating repetitive data processing tasks 
    \item Enhancing predictive modeling accuracy 
    \item Visualizing complex datasets more effectively 
    \item Supporting faster decision-making 
    \item Helping communicate findings to non-technical stakeholders 
\end{itemize}
\commentspace

\item[Q22.] What concerns or challenges do you face when integrating AI tools into your data science workflow? Select all that apply.
\resp{Complexity in setting up and using AI tools; Accuracy and reliability of AI predictions; Understanding how AI models generate outputs; Data security and privacy risks; AI’s impact on data science job roles.}
\commentspace

\begin{figure}[H]
    \centering
    \includegraphics[width=0.85\textwidth]{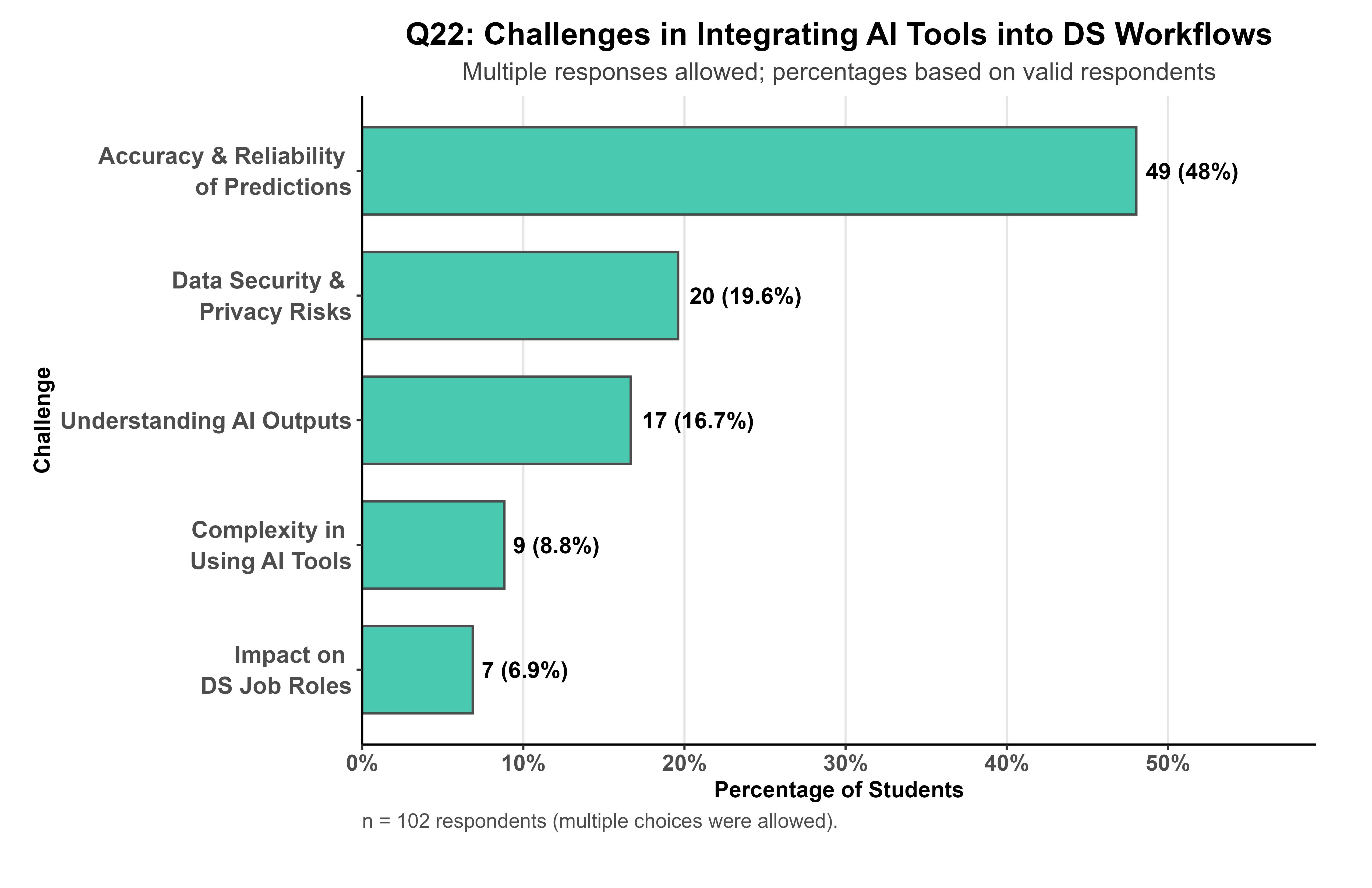}
    \caption{Challenges faced by students when integrating AI tools into  DS  workflow.}
    \label{fig:Q22-challenges}
\end{figure}

\item[Q23.] To what extent do you agree with the following statements regarding the use of AI tools in education in general?
\resp \LikertSA
\begin{itemize}
    \item Using generative AI technologies such as ChatGPT to complete assignments undermines the value of university education 
    \item Generative AI technologies such as ChatGPT may limit opportunities to interact with others and socialize while completing coursework 
    \item Generative AI technologies such as ChatGPT may hinder the development of transferable skills such as teamwork, problem-solving, and leadership 
    \item I may become over-reliant on generative AI technologies 
\end{itemize}
\commentspace
\begin{figure}[H]
    \centering
    \includegraphics[width=0.85\textwidth]{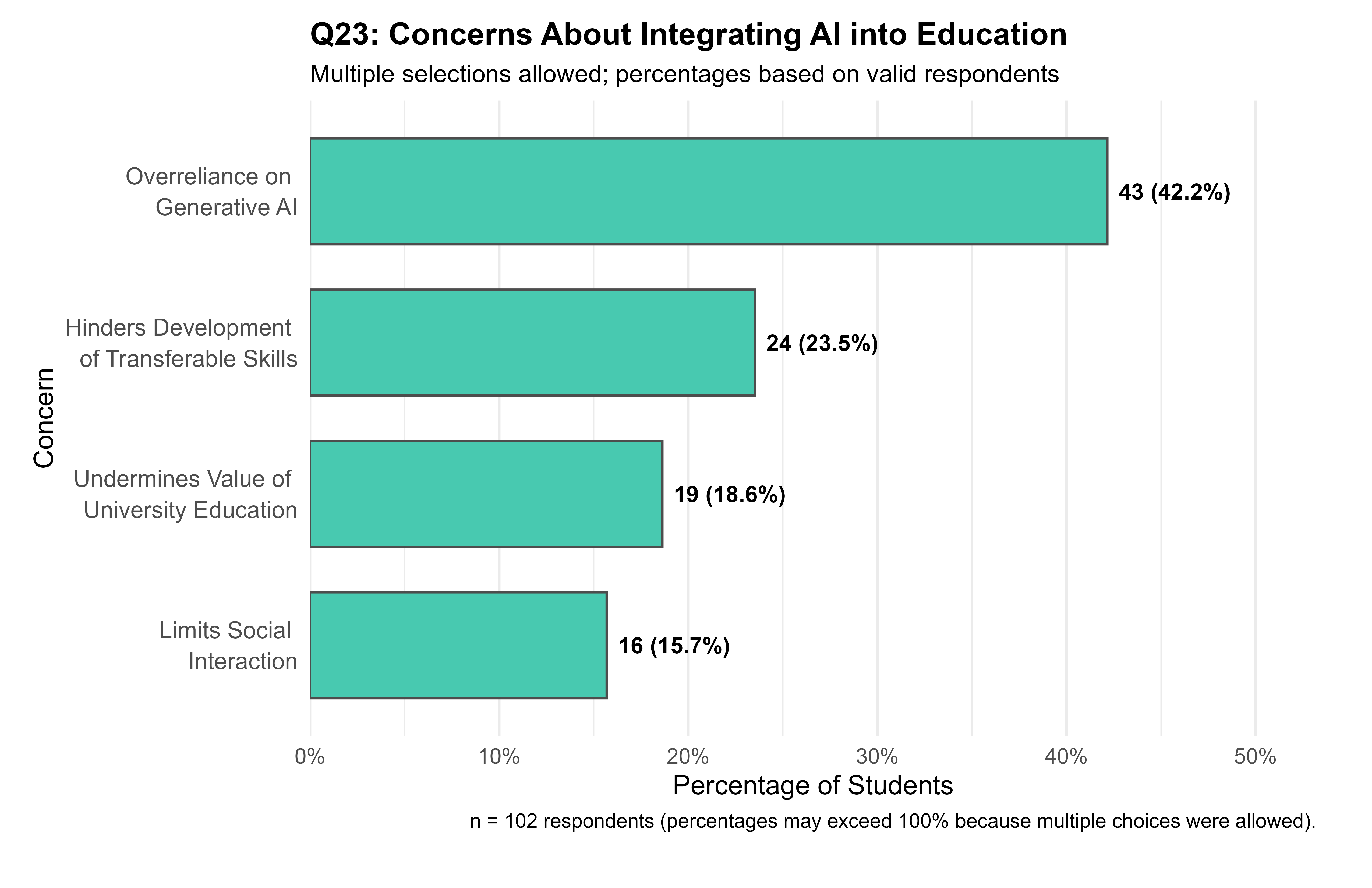}
   \caption{Students' concerns about AI use in education.}
    \label{fig:Studentconcerns}
\end{figure}

\item[Q24.] Do you believe that AI tools should complement traditional methods in data science or replace them?
\resp{AI tools do not complement traditional data science tools; AI tools complement traditional methods; Balance between AI tools and traditional methods; AI tools mostly replace traditional methods; Rely entirely on AI tools.}
\commentspace
\begin{figure}[H]
    \centering
    \includegraphics[width=0.85\textwidth]{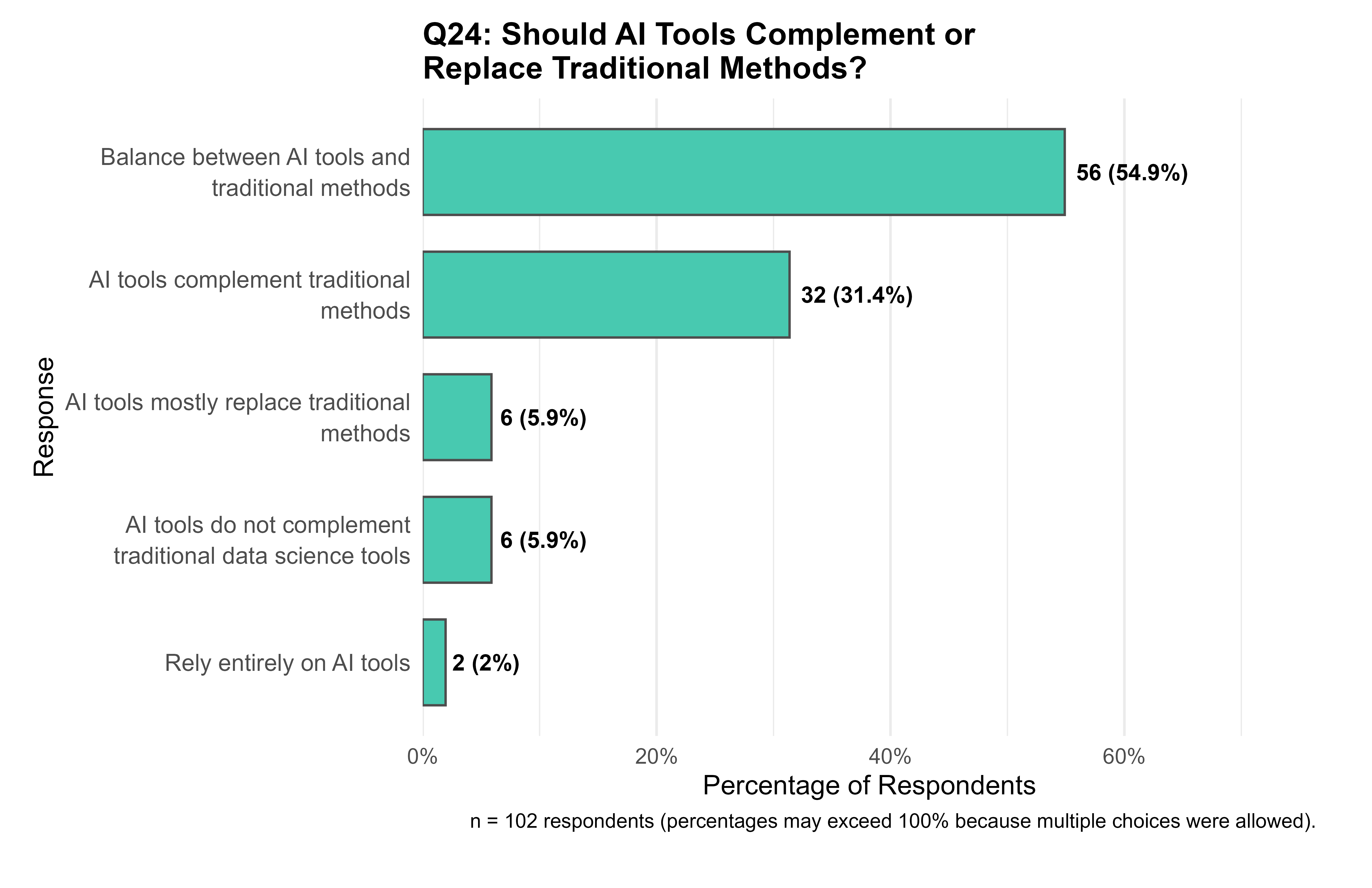}
    \caption{Students’ views on whether AI tools should complement or replace traditional methods.}
    \label{fig:Q24}
\end{figure}
\end{qenum}


\section{Faculty Survey Questions}\label{sec:faculty_survey}

\title{Faculty Survey Questions}
 This section includes detailed faculty survey questions, response options, and additional secondary results or visualizations.
\noindent\textit{Note. Questions Q1--Q2 were administrative consent items. Figures for questions whose primary results appear in the main text (Q9, Q10, Q11, Q13, Q15) are not duplicated here; see the corresponding main-text figures. Q16 was omitted during survey validation due to redundancy. All other response distributions are presented immediately following each question.}

\vspace{1em}


\setlist[qenum]{
  style=nextline,
  font=\normalfont\bfseries,
  leftmargin=2.4em,
  labelsep=0.6em,
  itemsep=1.15em,
  topsep=0.5em
}

\setlist[itemize]{
  leftmargin=1.8em,
  itemsep=0.25em,
  topsep=0.25em
}

\newcommand{\LikertConcern}{\emph{(Not concerned--Very concerned)}}
\newcommand{\LikertSkill}{\emph{(No experience--Expert)}}

\vspace{1em}

\begin{qenum}

\item[Q3.] What is your gender?
\resp{Male; Female; Non-binary; Trans man; Trans woman; Prefer not to say.}
\commentspace

\item[Q4.] What is your primary area of expertise within data science?
\resp{Open-ended response.}
\commentspace

\item[Q5.] How many years have you been teaching data science or related subjects?
\resp{1--3 years; 3--5 years; 5--10 years; 10+ years.}
\commentspace

\item[Q6.] At what academic level do you primarily teach?
\resp{Undergraduate; Graduate; Both.}
\commentspace

\item[Q7.] Have you incorporated AI tools in your data science teaching?
\resp{Never; Rarely; Sometimes; Often; Always.}

\begin{figure}[H]
    \centering
    \includegraphics[width=0.8\textwidth]{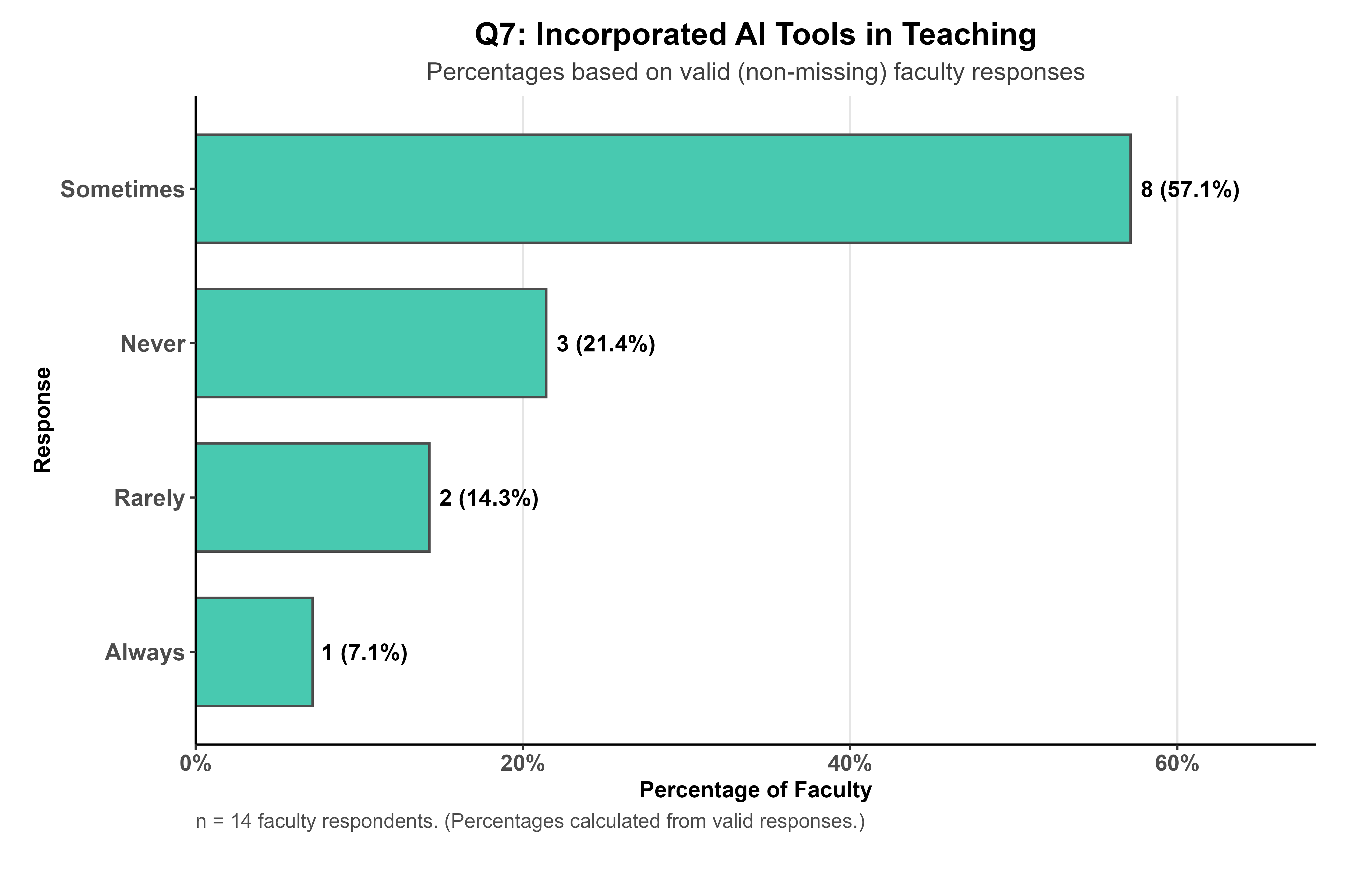}
    \caption{Faculty self-reported frequency of incorporating AI tools in teaching.}
    \label{fig:Q7_used_AI}
\end{figure}

{\noindent A majority of faculty (57.1\%) reported incorporating AI tools ``Sometimes,'' while 21.4\% had never done so and 7.1\% reported ``Always.'' See Section~S1.4 for further discussion.}
\commentspace

\item[Q8.] What AI tools or platforms do you use in your teaching? Select all that apply.
\resp{AI-driven coding platforms (e.g., Jupyter Lab, Colab); Automated grading systems; AI-driven learning assistants (e.g., ChatGPT); AI-based data analysis or visualization tools (e.g., DataRobot, Tableau); AI-based project management or collaboration tools; Other (please specify); None.}

{\noindent AI learning assistants such as ChatGPT were used by all 14 respondents (100\%), followed by AI-driven coding platforms (64.3\%) and data analysis or visualization tools (42.9\%). See Section~S1.4 for further discussion.}
\commentspace

\item[Q9.] How would you rate your ability to perform the following tasks using AI tools in your teaching?
\resp \LikertSkill
\begin{itemize}
    \item Implement AI-assisted coding environments for teaching 
    \item Use AI tools for real-time data analysis and visualization in class 
    \item Automate student feedback and assessment using AI tools 
    \item Guide students in applying AI to solve real-world data science problems 
    \item Employ AI for student performance tracking and curriculum adjustments 
\end{itemize}
\commentspace
{\noindent\textit{The response distribution for this question is presented in Figure~4 of the main text. Over 60\% of faculty rated their skill level as ``Beginner'' or ``No Experience'' across all tasks. See Section~S1.4 for further discussion.}}

\item[Q10.] How frequently do you use AI tools in the following teaching activities?
\resp \LikertFreq
\begin{itemize}
    \item Automating grading or feedback for data science assignments 
    \item Engaging students through AI-assisted interactive labs (e.g., coding or machine learning exercises) 
    \item Monitoring student learning progress using AI-based analytics 
    \item Generating teaching materials with AI support 
\end{itemize}
\commentspace
{\noindent\textit{The response distribution for this question is presented in Figure~5 of the main text. Fewer than 23\% of faculty reported frequent use for any activity. See Section~S1.4 for further discussion.}}

\item[Q11.] To what extent do you agree with the following statements about AI in data science education?
\resp{\LikertSA}
\begin{itemize}
    \item AI tools are vital for future data science education 
    \item AI can significantly improve the teaching of data science topics such as machine learning, deep learning, and big data 
    \item Overreliance on AI tools may reduce the development of critical thinking skills in students 
    \item AI tools can efficiently automate repetitive teaching tasks, freeing up time for more complex educational engagement 
    \item AI tools are capable of improving equity and inclusion in data science education 
\end{itemize}
\commentspace
{\noindent\textit{The response distribution for this question is presented below. Over 75\% agreed on equity/inclusion and instruction in complex topics; concerns about critical thinking showed more ambivalence. See Section~S1.5 for further discussion.}}
 \begin{figure}[H]
    \centering
    \includegraphics[width=0.85\textwidth]{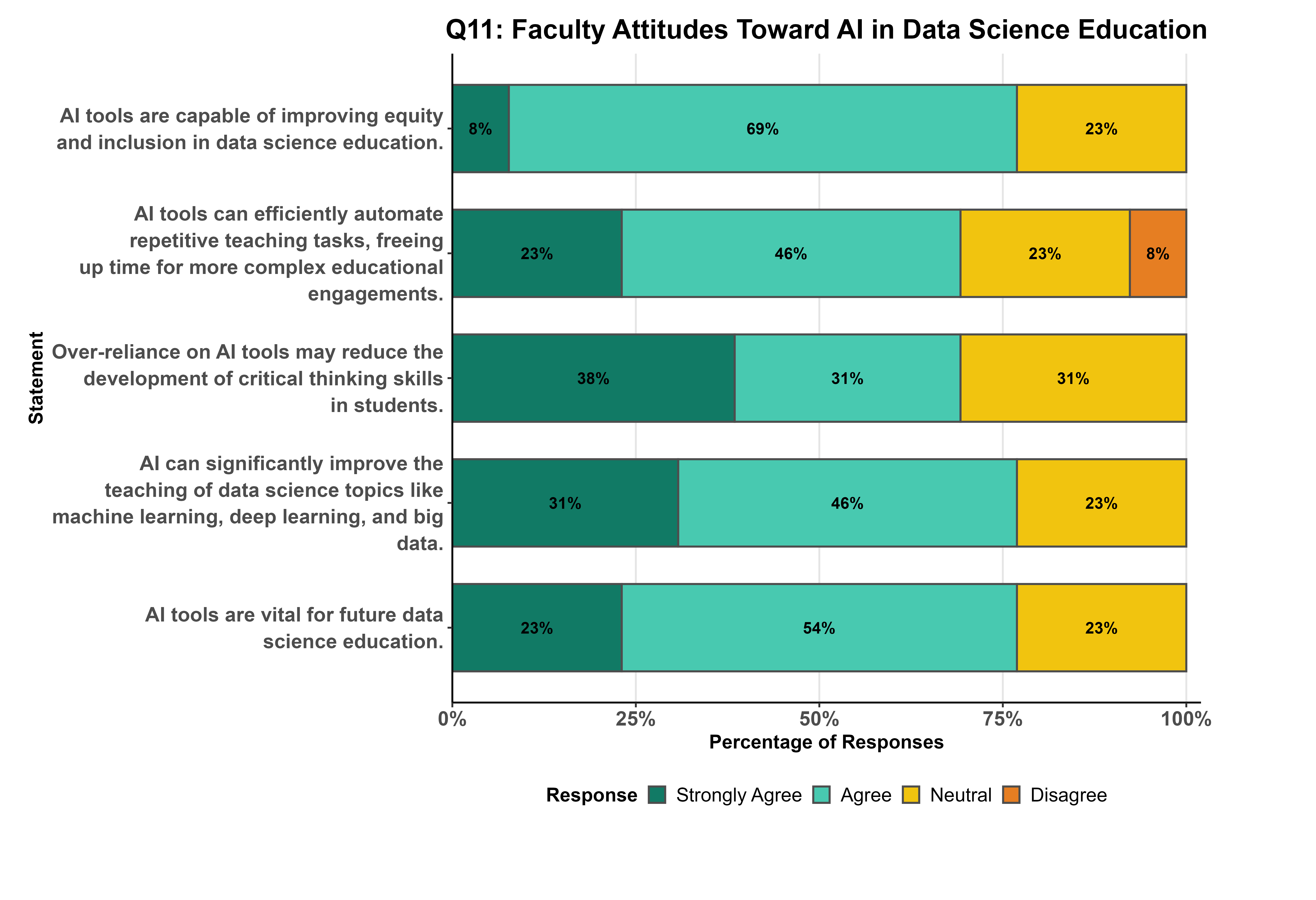}
    \caption{Faculty perceptions of the potential of AI tools to improve DS education.}
    \label{fig:Fac_attitudes}
\end{figure}

\item[Q12.] Do you believe that AI tools should complement or replace traditional methods of teaching data science?
\resp{AI tools should complement traditional methods; AI tools should balance with traditional methods; AI tools should replace traditional methods in some contexts; AI tools should not replace traditional methods.}

\begin{figure}[H]
    \centering
    \includegraphics[width=0.8\textwidth]{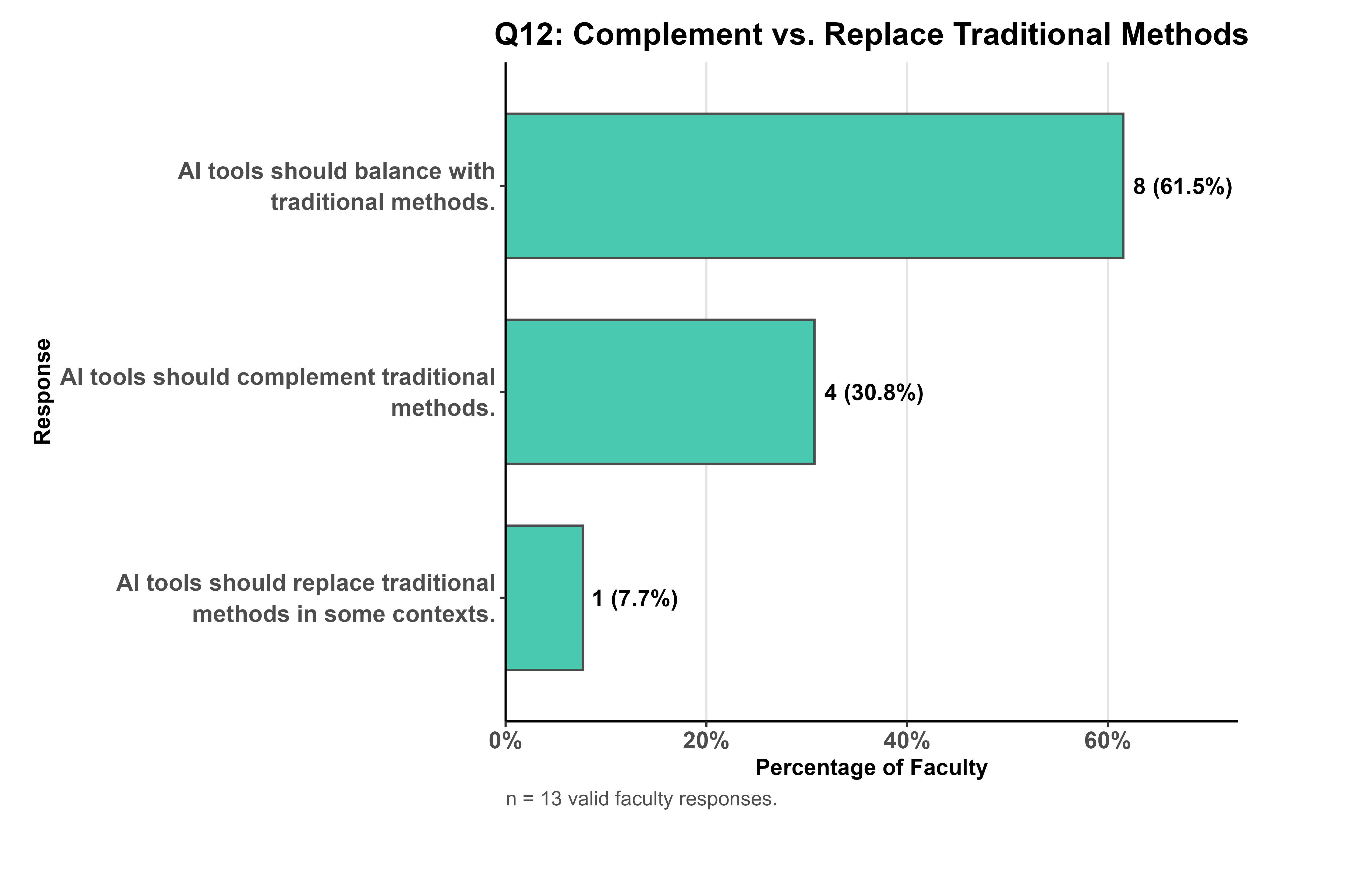}
    \caption{Faculty perspectives: AI as complement vs.\ replacement in education.}
    \label{fig:Q12_Complement_vs_Replace}
\end{figure}

{\noindent A majority (61.5\%) favored balancing AI with traditional methods, 30.8\% preferred complementary use, and only 7.7\% supported replacement in some contexts. No faculty endorsed full replacement. See Section~S1.5 for further discussion.}
\commentspace

\item[Q13.] How concerned are you about the following aspects of AI integration in data science education?
\resp{\LikertConcern}
\begin{itemize}
    \item Potential for AI to produce biased results in student assessments 
    \item Lack of transparency in how AI tools generate outputs (e.g., black-box models) 
    \item The challenge of keeping up with evolving AI technologies in education 
    \item Over-reliance on AI tools, leading to reduced faculty engagement with students 
    \item Data privacy and security issues related to using AI in education 
\end{itemize}

{\noindent\textit{The response distribution for this question is presented in Figure~8 of the main text. Faculty were unanimous in their concern about keeping up with evolving technologies. See Section~S2.6 for further discussion.}}
\commentspace

\item[Q14.] What are the biggest challenges you face in integrating AI tools into your data science courses? Select all that apply.
\resp{Difficulty in learning and mastering AI tools; Insufficient institutional support or training; Ethical concerns (e.g., fairness, bias); Limited access to advanced AI tools or platforms; Resistance from students or faculty to adopt AI tools; Balancing AI tools with the curriculum's learning objectives.}

\begin{figure}[H]
    \centering
    \includegraphics[width=0.8\textwidth]{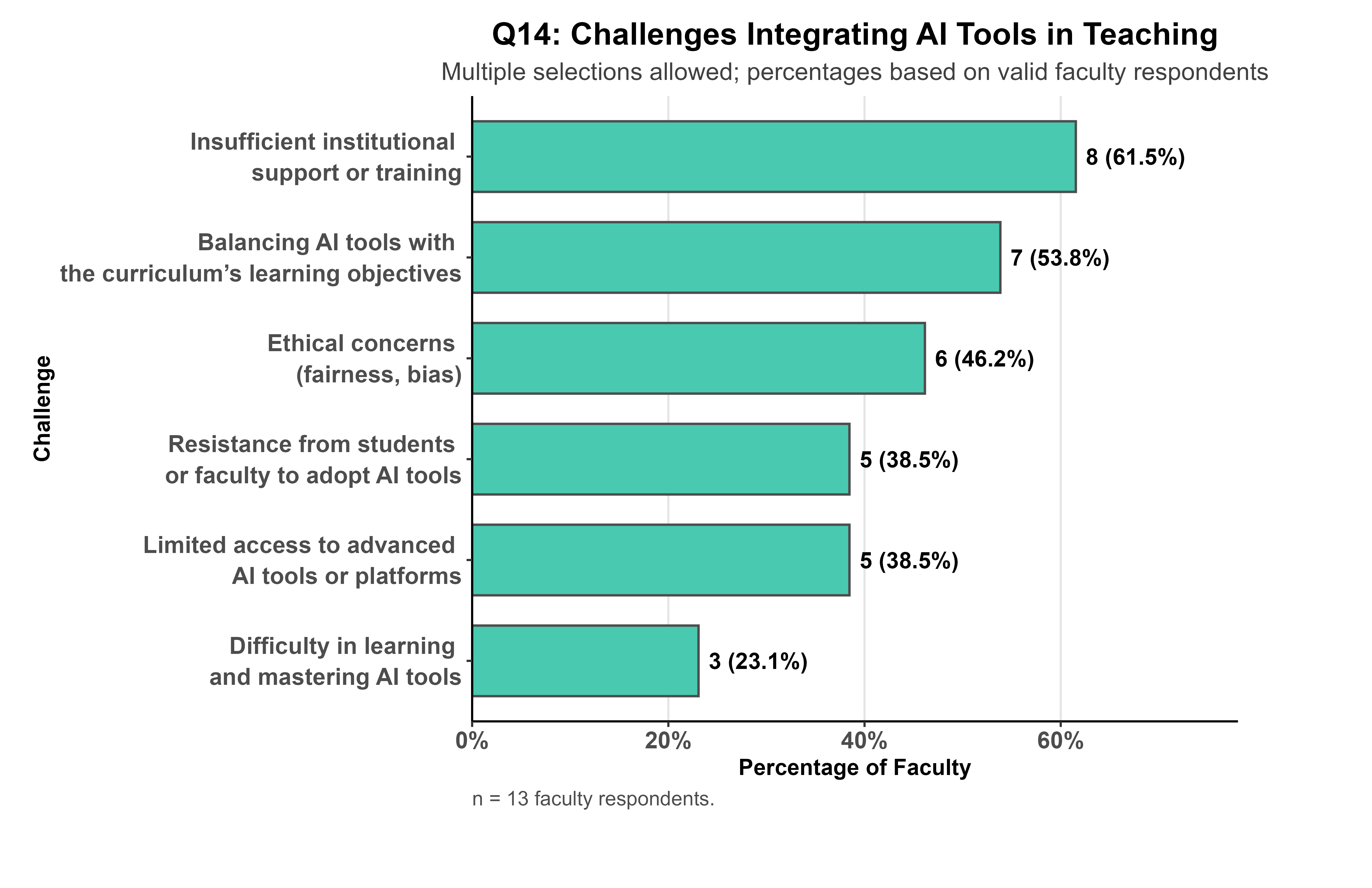}
    \caption{Reported challenges related to AI use in teaching.}
    \label{fig:Q14_Challenges_AI}
\end{figure}

{\noindent Insufficient institutional support (61.5\%) and balancing AI with curriculum objectives (53.8\%) were the most frequently cited challenges. See Section~S2.6 for further discussion.}
\commentspace

\item[Q15.] To what extent do you agree that AI tools can enhance the following aspects of data science teaching?
\resp{\LikertSA}
\begin{itemize}
    \item Automating assessment and feedback for data science projects 
    \item Supporting the personalization of learning paths based on student performance 
    \item Simplifying complex data science concepts through AI-enhanced visualizations
    \item Assisting with real-time performance tracking and identifying areas where students struggle 
    \item Enhancing student engagement with AI-driven interactive learning modules 
\end{itemize}

{\noindent\textit{The response distribution for this question is presented in Figure~7 of the main text. Strongest endorsement was for AI-enhanced visualizations; perceptions were most mixed for personalizing learning paths. See Section~S1.5 for further discussion.}}
\commentspace

\item[Q17.] How aware are you of your institution's guidelines on the use of AI in teaching and assessments?
\resp{Very aware; Somewhat aware; Neutral; Not aware.}

\begin{figure}[H]
    \centering
    \includegraphics[width=0.8\textwidth]{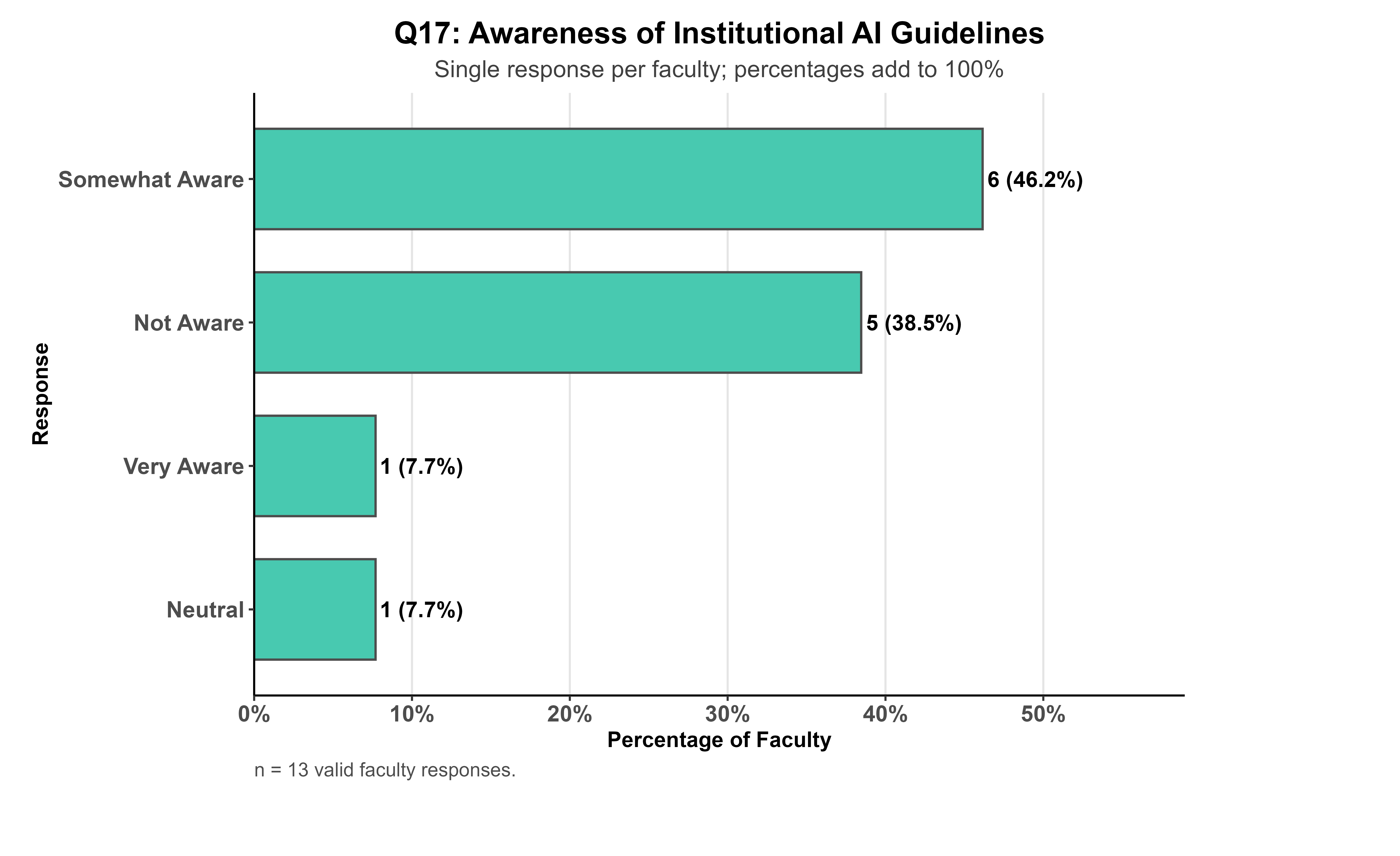}
    \caption{Faculty awareness of institutional AI guidelines.}
    \label{fig:Q17_faculty_awareness_AI_guidelines}
\end{figure}

{\noindent Only 7.7\% reported being ``Very Aware'' of institutional guidelines; 38.5\% were ``Not Aware.'' See Section~S1.6 for further discussion.}
\commentspace

\item[Q18.] What primary ethical concern do you have about using AI tools in education?
\resp{Bias in AI-driven student evaluations; Infringement on student privacy with AI monitoring tools; over-reliance on AI leading to less human oversight in teaching and assessments.}

\begin{figure}[H]
    \centering
    \includegraphics[width=0.8\textwidth]{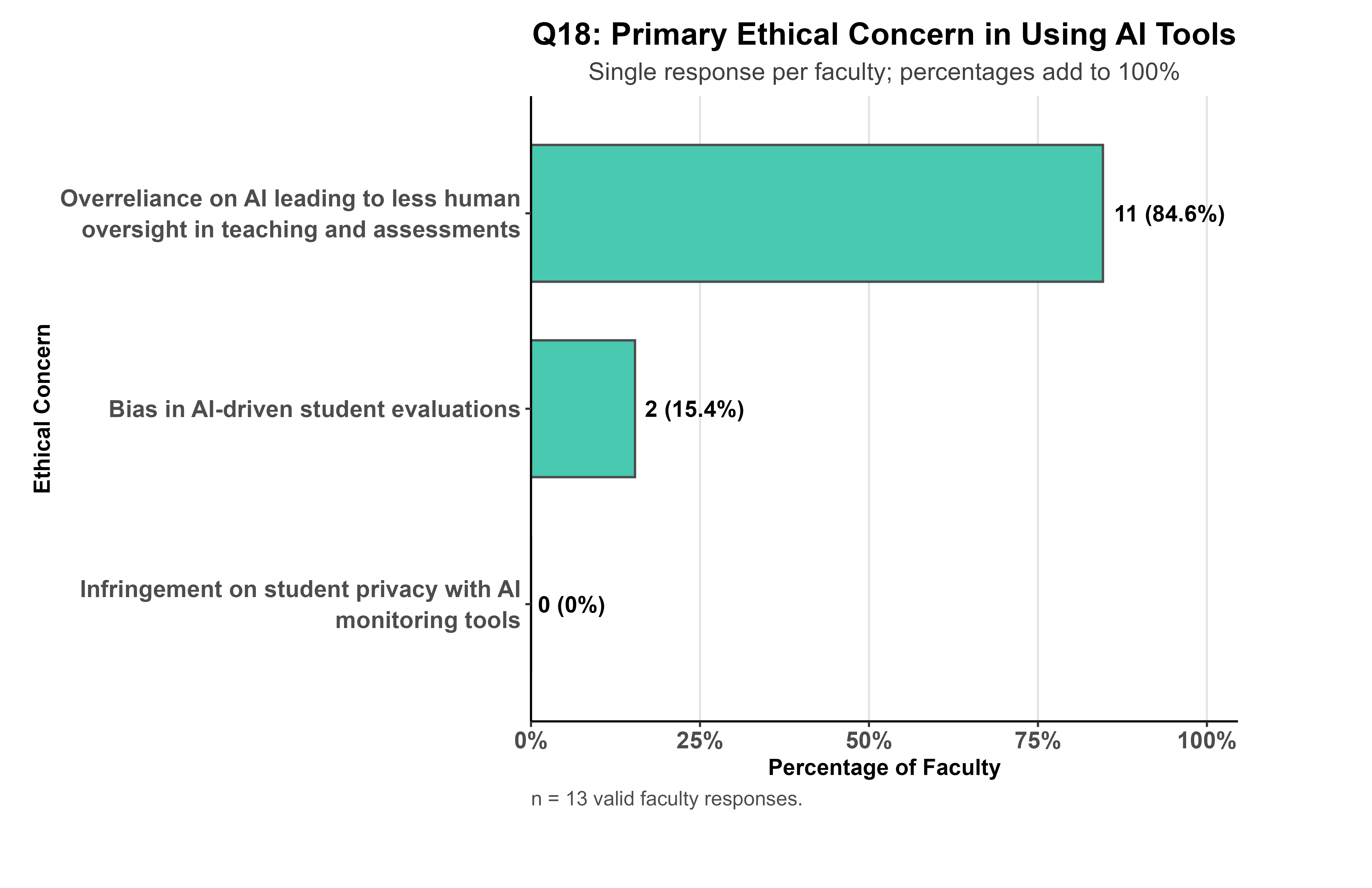}
    \caption{Primary ethical concerns in AI use.}
    \label{fig:Q18_Primary_Ethical_Concern}
\end{figure}

{\noindent Over-reliance on AI leading to less human oversight was cited by 84.6\% of faculty; 15.4\% cited bias in AI-driven evaluations. See Section~S1.6 for further discussion.}
\commentspace

\item[Q19.] How important is it to maintain human oversight when using AI for assessments and student feedback?
\resp{Very important; Important; Neutral; Not important.}

\begin{figure}[H]
    \centering
    \includegraphics[width=0.8\textwidth]{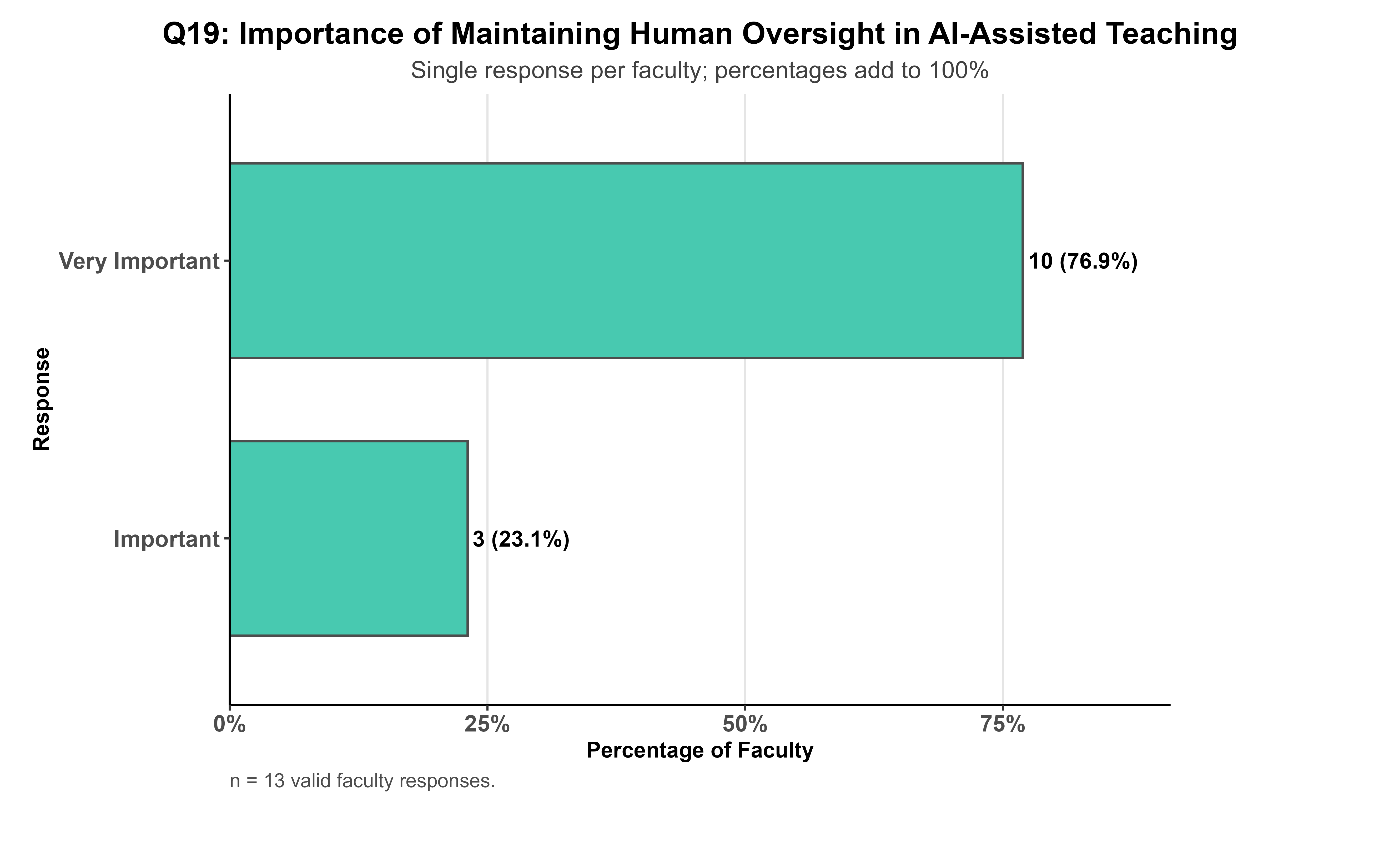}
    \caption{Faculty perspectives on human oversight in AI applications.}
    \label{fig:Q19_Human_Oversight}
\end{figure}

{\noindent Faculty unanimously endorsed the importance of human oversight: 76.9\% rated it ``Very Important'' and 23.1\% ``Important.'' See Section~S1.6 for further discussion.}
\commentspace

\item[Q20.] Which of the following best describes the AI policy in your syllabus/class?
\resp
\begin{itemize}
    \item Assistance from generative AI technologies is not allowed for any activities or assignments in this course; 
     \item The use of generative AI technologies is allowed for specific activities or assignments designated by the instructor, and students must clearly indicate when AI-generated content has been used and cite the specific model and platform employed; 
      \item The use of generative AI technologies is permitted for activities and assignments in this course, and proper attribution is required whenever AI-generated content is used.
\end{itemize}

\begin{figure}[H]
    \centering
    \includegraphics[width=0.8\textwidth]{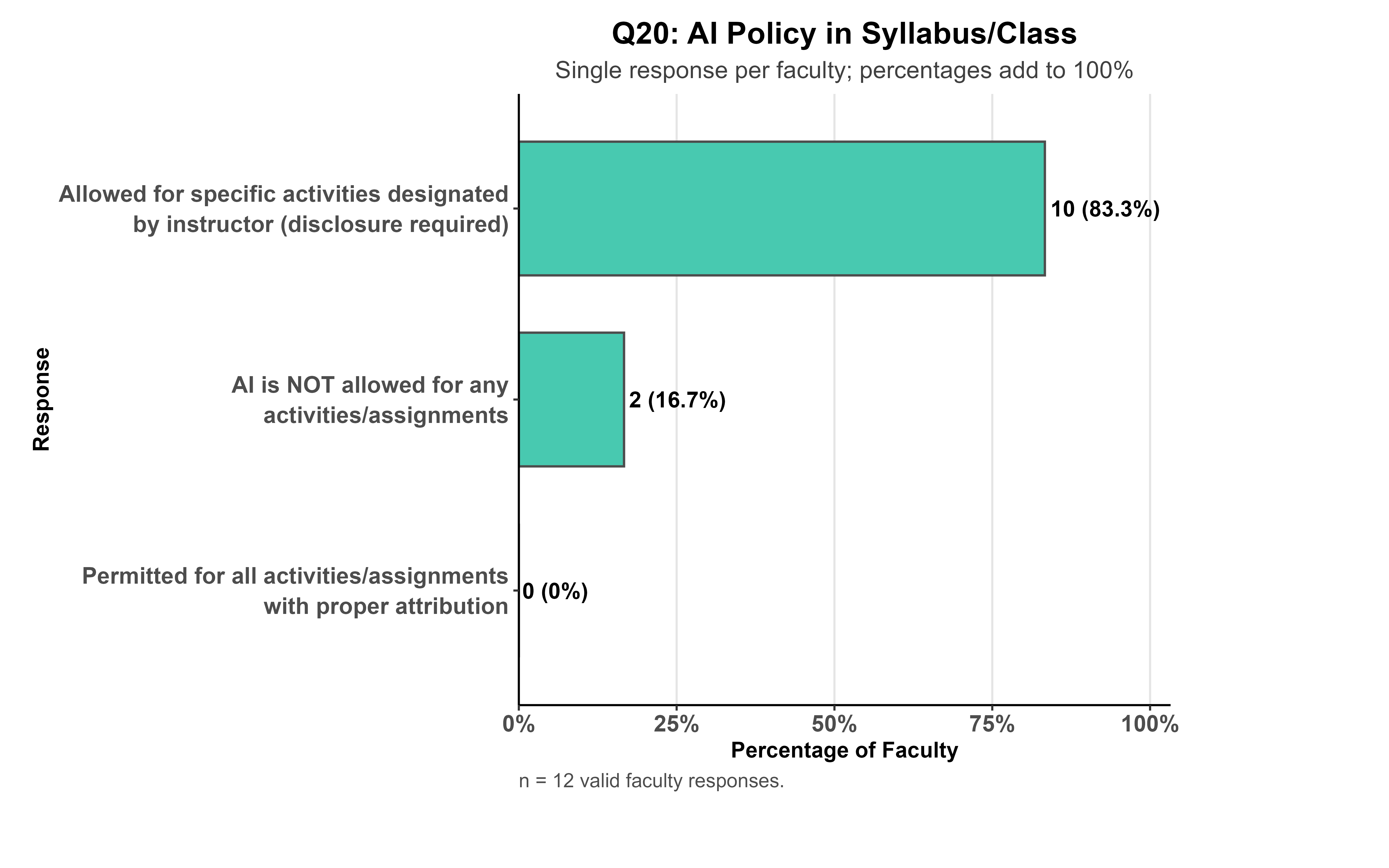}
    \caption{AI policy statements in course syllabi/ class.}
    \label{fig:Q20_AI_Policy_Syllabus}
\end{figure}

{\noindent The majority (83.3\%) allowed AI for instructor-designated activities with disclosure required; 16.7\% prohibited AI entirely; none permitted unrestricted use. See Section~S1.6 for further discussion.}
\commentspace

\item[Q21.] How confident are you in your knowledge of AI tools available for data science education?
\resp{Very confident; Somewhat confident; Neutral; Not confident.}

\begin{figure}[H]
    \centering
    \includegraphics[width=0.8\textwidth]{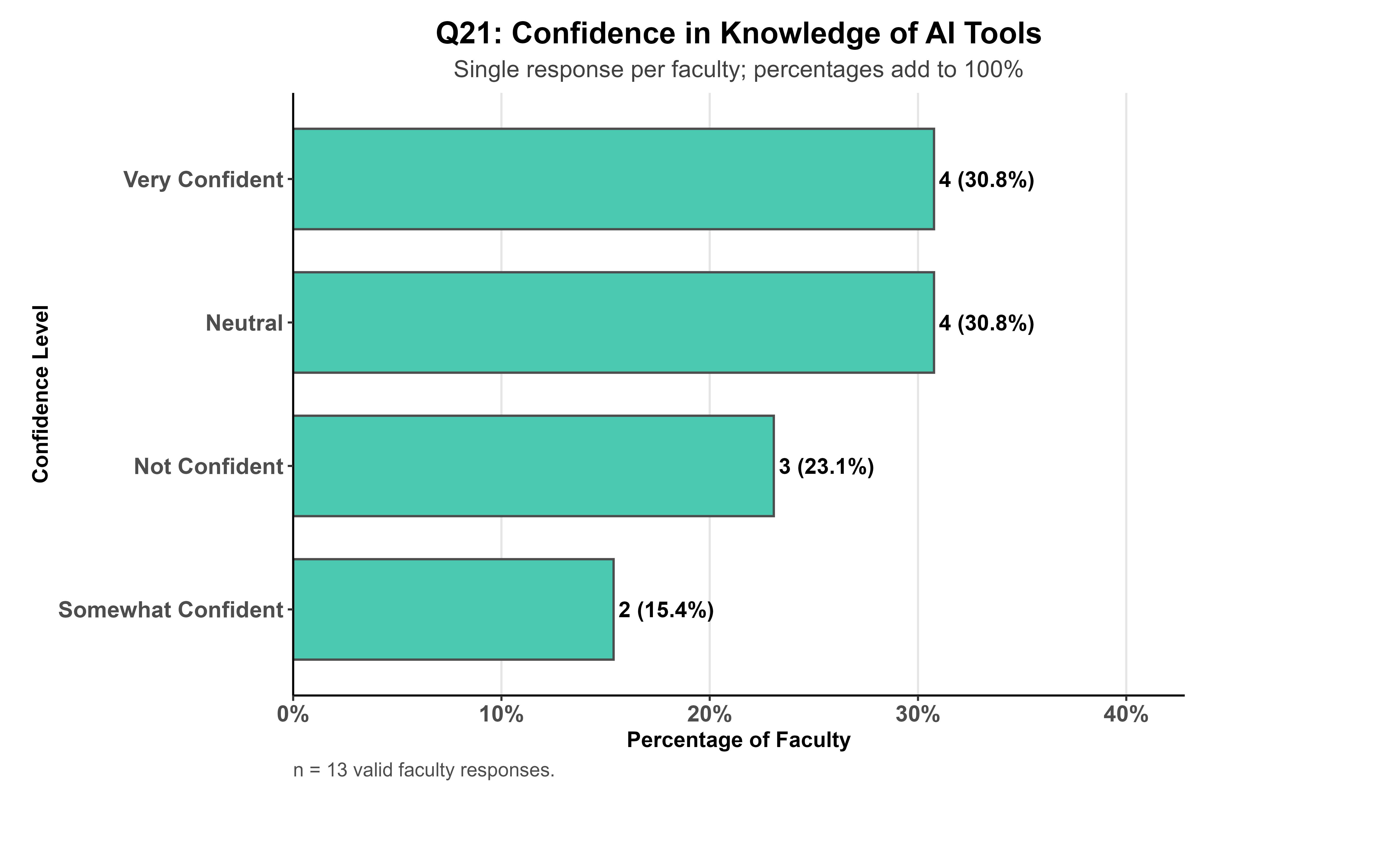}
    \caption{Faculty confidence in using AI tools.}
    \label{fig:Q21_confidence_in_AI_tools}
\end{figure}

{\noindent Responses were distributed across all levels: 30.8\% ``Very Confident,'' 15.4\% ``Somewhat Confident,'' 30.8\% ``Neutral,'' and 23.1\% ``Not Confident.'' See Section~S1.6 for further discussion.}

\item[Q22.] What types of training or support would you need to effectively integrate AI tools into your teaching? Select all that apply.
\resp{More AI-related training or professional development; Access to advanced AI platforms for classroom use; Technical support for integrating AI into the curriculum; Institutional support for purchasing AI tools; Peer collaboration and sharing best practices.}

\begin{figure}[H]
    \centering
    \includegraphics[width=0.85\textwidth]{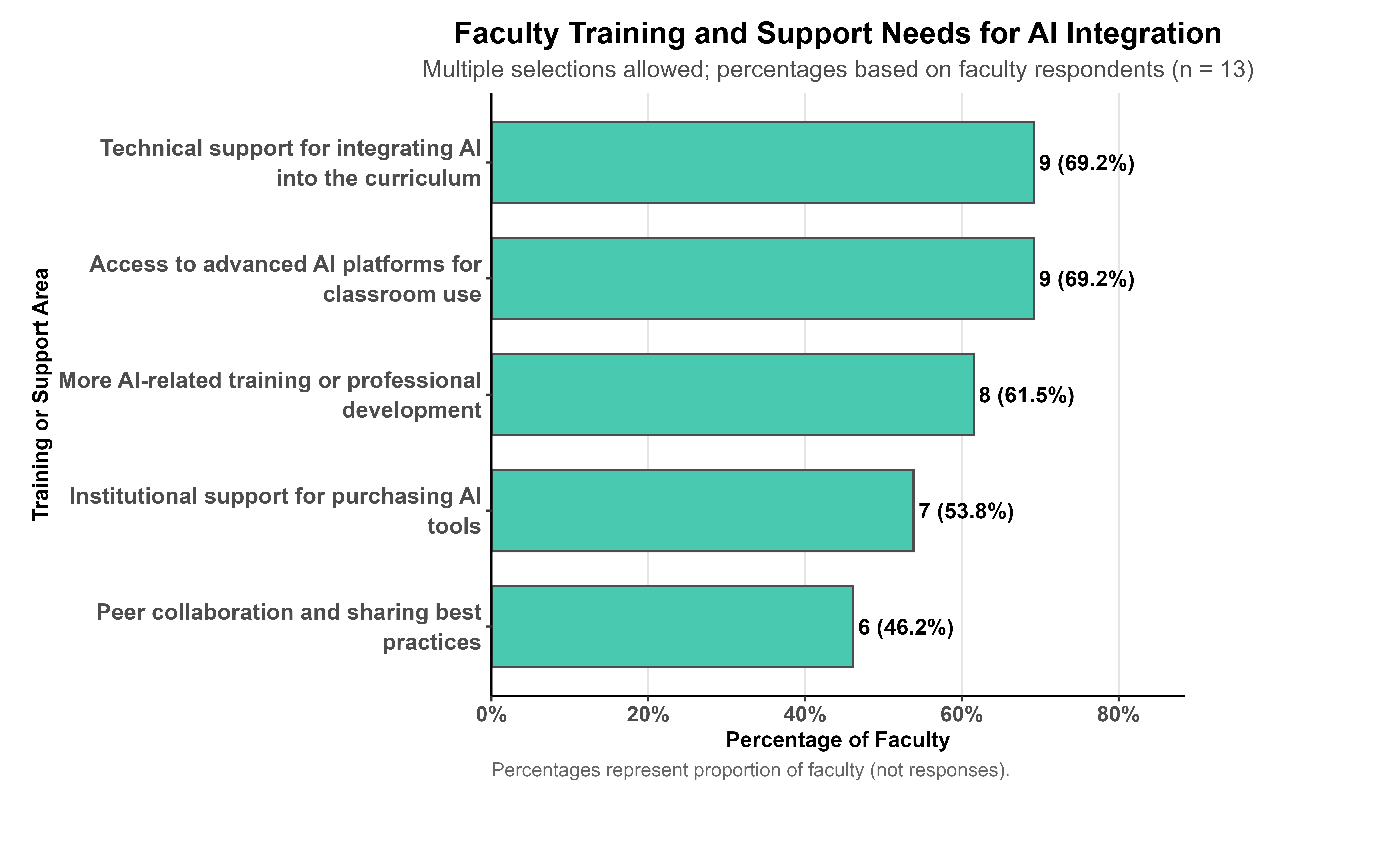}
    \caption{Faculty reported training needs to integrate AI into their curricula.}
    \label{fig:Fac-training-Needs_Q22}
\end{figure}

{\noindent Access to AI platforms and technical integration support were each cited by 69.2\%; AI-related professional development by 61.5\%. See Section~S1.6 for further discussion.}
\commentspace

\item[Q23.] Do you believe your institution provides sufficient guidance and resources for using AI in teaching?
\resp{Yes; No; Somewhat.}

\begin{figure}[H]
    \centering
    \includegraphics[width=0.8\textwidth]{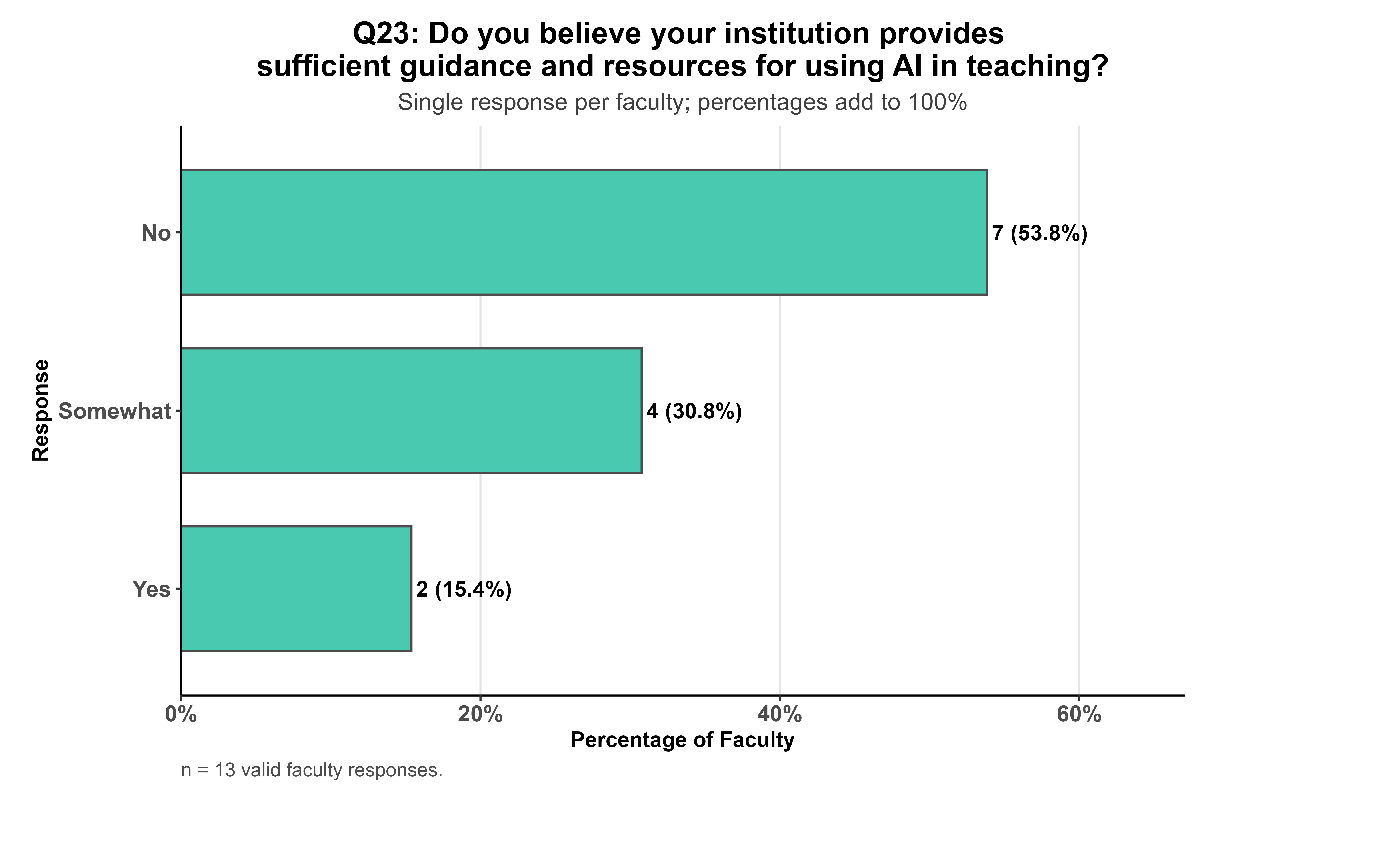}
    \caption{Do faculty believe their institution provides sufficient guidance and resources for using AI in teaching?}
    \label{fig:Q23_Institutional_Guidance}
\end{figure}

{\noindent A majority (53.8\%) responded ``No,'' 30.8\% ``Somewhat,'' and only 15.4\% ``Yes.'' See Section~S1.6 for further discussion.}

\end{qenum}